\documentclass{pasa}

\usepackage{color}
\usepackage{lscape}
\usepackage{deluxetable}

\newcommand{\snia}{SN~Ia}
\newcommand{\sneia}{SNe~Ia}

\newcommand{\wifes}{WiFeS}
\newcommand{\pywifes}{PyWiFeS}

\newcommand{\nobj}{175}
\newcommand{\nspec}{357}
\newcommand{\nnights}{82}

\newcommand{\arcsec}{$''$}
\newcommand{\kms}{km\,s$^{-1}$}

\title[The AWSNAP Program]{The ANU WiFeS SuperNovA Program (AWSNAP)}
\author[Childress et al.]{
Michael~J.~Childress$^{1,2,3}$\thanks{E-mail: m.j.childress@soton.ac.uk},
Brad~E.~Tucker$^{1,2}$,
Fang~Yuan$^{1,2}$, 
Richard~Scalzo$^{1}$,
Ashley~Ruiter$^{1,2}$,
Ivo~Seitenzahl$^{1,2}$, 
Bonnie~Zhang$^{1}$, 
Brian~Schmidt$^{1}$,
Borja~Anguiano$^{4}$,
Suryashree~Aniyan$^{1}$,
Daniel~D.~R.~Bayliss$^{1,5}$,
Joao~Bento$^{1}$,
Michael~Bessell$^{1}$,
Fuyan~Bian$^{1}$,
Rebecca~Davies$^{1}$,
Michael~Dopita$^{1}$,
Lisa~Fogarty$^{6}$,
Amelia~Fraser-McKelvie$^{7,8}$,
Ken~Freeman$^{1}$,
Rajika~Kuruwita$^{1}$,
Anne~M.~Medling$^{1}$,
Simon~J.~Murphy$^{1}$,
Simon~J.~Murphy$^{6,9}$,
Matthew~Owers$^{4,10}$,
Fiona~Panther$^{1,2}$,
Sarah~M.~Sweet$^{1}$,
Adam~D.~Thomas$^{1}$,
George~Zhou$^{11,1}$
\\
\affil{$^1$Research School of Astronomy and Astrophysics,
  Australian National University,
  Canberra, ACT 2611, Australia}%
\affil{$^2$ARC Centre of Excellence for All-sky Astrophysics (CAASTRO)}%
\affil{$^3$School of Physics and Astronomy,
  University of Southampton,
  Southampton, SO17 1BJ, UK}%
\affil{$^{4}$Department of Physics and Astronomy,
  Macquarie University,
  NSW 2109, Australia}
\affil{$^{5}$Observatoire Astronomique de l'Universit\'e de Gen\`eve,
  51 ch. des Maillettes,
  1290 Versoix, Switzerland}
\affil{$^{6}$Sydney Institute for Astronomy (SIfA),
  School of Physics,
  The University of Sydney, NSW 2006, Australia}
\affil{$^{7}$School of Physics and Astronomy,
  Monash University,
  Clayton, Victoria 3800, Australia}
\affil{$^{8}$Monash Centre for Astrophysics (MoCA),
  Monash University,
  Clayton, Victoria 3800, Australia}
\affil{$^{9}$Stellar Astrophysics Centre,
  Department of Physics and Astronomy,
  Aarhus University,
  8000 Aarhus C, Denmark}
\affil{$^{10}$ Australian Astronomical Observatory,
  PO Box 915,
  North Ryde, NSW 1670, Australia}
\affil{$^{11}$Harvard-Smithsonian Center for Astrophysics,
  60 Garden St,
  Cambridge, MA 02138, USA}
}
\jid{PASA}
\doi{10.1017/pas.\the\year.xxx}
\jyear{\the\year}

\usepackage[authoryear]{natbib}
\bibpunct{(}{)}{;}{a}{}{,}
\begin{document}%

\begin{abstract}
This paper presents the first major data release and survey description for the ANU WiFeS SuperNovA Program (AWSNAP).  AWSNAP is an ongoing supernova spectroscopy campaign utilising the Wide Field Spectrograph (\wifes) on the Australian National University (ANU) 2.3m telescope.
The first and primary data release of this program (AWSNAP-DR1) releases \nspec\ spectra of \nobj\ unique objects collected over \nnights\ equivalent full nights of observing from July 2012 to August 2015.  These spectra have been made publicly available via the WISEREP supernova spectroscopy repository.

We analyse the AWSNAP sample of Type Ia supernova spectra, including measurements of narrow sodium absorption features afforded by the high spectral resolution of the \wifes\ instrument.  In some cases we were able to use the integral-field nature of the \wifes\ instrument to measure the rotation velocity of the SN host galaxy near the SN location in order to obtain precision sodium absorption velocities.  We also present an extensive time series of SN~2012dn, including a near-nebular spectrum which both confirms its ``super-Chadrasekhar'' status and enables measurement of the sub-solar host metallicity at the SN site.

\end{abstract}
\maketitle%
%

\section{Introduction}
In the last decade, wide-field extragalactic transient surveys -- such as the Palomar Transient Factory \citep[PTF;][]{rau09, law09}, the Panoramic Survey Telescope and Rapid Response System \citep[PanSTARRS;][]{panstarrs}, the Catalina Real-time Transient Survey \citep[CRTS;][]{crts}, and the Texas Supernova Search \citep{quimbythesis, yuanthesis} -- have revolutionised our understanding of the myriad ways in which stars explode through the discovery of new classes of exotic transients.  Simultaneously, these surveys have discovered hundreds of supernovae (SNe) of ``traditional'' types \citep[see][for a review]{filippenko97}, enabling statistical analyses of the properties of these SNe.

While imaging surveys have provided discovery and light curves for this wealth of new transients, complementary spectroscopy surveys have provided the critical insight into the physical origins of these events.  Numerous supernova spectroscopy surveys have released thousands of high quality spectra of nearby SNe into the public domain \citep{matheson08, blondin12, bsnip1, folatelli13, modjaz14}.  These surveys have frequently been dedicated to the spectroscopic followup of Type Ia supernovae (\sneia) which, due to their rates and luminosities, dominate any magnitude-limited imaging survey.  Such surveys have revealed that photometrically similar SNe can still exhibit diversity of spectroscopic behaviour, indicating spectra remain a critical tool for revealing the full nature of the supernova progenitors (particularly for \sneia).  Additionally, spectra remain critical for supernova classification -- particularly at early phases when the full photometric evolution has yet to be revealed.  Such early classifications then inform the use of additional SN followup facilities, including those operating outside the optical window.

Recently the Public ESO Spectroscopic Survey for Transient Objects \citep[PESSTO;][]{pessto} began a multi-year program on the NTT 3.6m telescope in Chile, with the goal of obtaining high quality spectral time series for roughly 100 transients (of all kinds) to be released to the public.  This survey has already released hundreds of spectra in its first two annual data releases, and continues to release all SN classificaion spectra within typically 1 day from observation.  Other ongoing SN spectroscopy programs, such as the Asiago Supernova Program \citep{tomasella14}, also make important contributions to the transient community through timely SN classification and spectroscopy releases.

Here we describe our ongoing spectroscopy program AWSNAP -- the {\bf A}NU {\bf W}iFeS {\bf S}uper{\bf N}ov{\bf A} {\bf P}rogram -- which uses the Wide Field Spectrograph \citep[WiFeS;][]{dopita07, dopita10} on the Australian National University (ANU) 2.3m telescope at Siding Spring Observatory in Australia.  In this paper we describe the data processing procedures for this ongoing program, and describe the first AWSNAP data release (AWSNAP DR1) comprising \nspec\ spectra of \nobj\ supernova of various types obtained during \nnights\ classically-scheduled observing nights over a 3 year period from July 2012 to August 2015.  Most of these spectra have been released publically via the Weizmann Interactive Supernova data REPository \citep[WISeREP\footnote{http://wiserep.weizmann.ac.il} --][]{wiserep}, with the remainder set to be released within the next year as part of forthcoming PESSTO papers.  This program will continue to observe SNe of interest and classify SN discoveries from transient searches such as the new SkyMapper Transients Survey \citep{keller07}.  We aim to release future SN classification spectra from AWSNAP publicly via WISeREP in parallel with any classification announcements.

This paper is organised as follows. Section~\ref{sec:data} describes the \wifes\ data processing and SN spectrum extraction procedures.  Section~\ref{sec:sn_sample} presents general properties of our SN sample and compares AWSNAP DR1 to other public SN spectra releases.  In Section~\ref{sec:results} we present some analysis of the properties of the \sneia\ in our sample, including measurement of narrow sodium absorption features afforded by the intermediate resolution of the \wifes\ spectrograph.  Some concluding remarks follow in Section~\ref{sec:conclusions}.

\section{Observations and Data Description}
\label{sec:data}
Observations for AWSNAP were conducted with the Wide Field Spectrograph \citep[WiFeS --][]{dopita07, dopita10} on the Australian National University 2.3m telescope at Siding Spring Observatory in northern New South Wales, Australia.  Observing nights were classically scheduled with a single night of observing every 8-15 days.  On some occasions, special objects of interest were observed during non-AWSNAP nights.  A full list of the AWSNAP transient spectra is presented in Table~\ref{tab:all_awsnap_spectra} in Appendix~A.  In the sections that follow, we describe the processing of the \wifes\ data, then characterise both the long term performance of the \wifes\ instrument and observing conditions at Siding Spring.

\subsection{Data Reduction and Supernova Spectrum Extraction}
\label{sec:reduction}
The \wifes\ instrument is an image-slicing integral field spectrograph with a wide 25\arcsec$\times$38\arcsec\ field of view.  For AWSNAP, this frequently provided simultaneous integral field observations of SNe and their host galaxies.  The \wifes\ image slicer breaks the field of view into 25 ``slitlets'' of width 1\arcsec, which then pass through a dichroic beamsplitter and volume phase holographic (VPH) gratings before arriving at 4k $\times$ 4k CCD detectors.  AWSNAP observations were always conducted with a CCD binning of 2 in the vertical direction -- this sets the vertical spatial scale of the detector to be 1 \arcsec, yielding final integral field elements (or ``spaxels'') of size  1\arcsec$\times$1\arcsec.  Typically seeing at Siding Spring is roughly 2\arcsec\ (see Section~\ref{sec:sso_stats}).

The VPH gratings utilised by \wifes\ provide a higher wavelength resolution than traditional glass gratings.  The low- and high-resolution gratings provide resolutions of $R=3000$ and $R=7000$, respectively, yielding velocity resolutions of up to $\sigma_v \sim 45$~\kms\ which is ideal for observing nebular emission lines from ionised regions in galaxies.  For supernovae, this can reveal narrow absorption features (see Section~\ref{sec:sodium}) from circum-stellar material (CSM) which are typically smeared out by lower resolution spectrographs.  AWSNAP observations were generally conducted with the lower resolution B3000 and R3000 gratings for the blue and red arms of the spectrograph, respectively, with the RT560 dichroic beamsplitter.  Occasionally the R7000 grating was deployed on the red arm to provide higher resolution observations of sodium absorption features.  In Table~\ref{tab:wifes_gratings} we provide the wavelength range, spectroscopic pixel size, and wavelength resolution (determined as the FWHM of calibration lamp emission lines) for the three gratings used for AWSNAP observations.

\begin{table}
  \caption{Details of WiFeS gratings.}
  \label{tab:wifes_gratings}
\begin{tabular}{lcccc}
\hline
Grating & $\lambda_{min}$ & $\lambda_{max}$ & Pixel size & Resolution\\
\hline
B3000 & 3500 \AA\ & 5700 \AA\ & 0.77 \AA\ & 1.5 \AA\ \\
R3000 & 5400 \AA\ & 9570 \AA\ & 1.25 \AA\ & 2.5 \AA\ \\
R7000 & 5400 \AA\ & 7020 \AA\ & 0.44 \AA\ & 0.9 \AA\ \\
\hline
\end{tabular}
\end{table}

Data for AWSNAP observations were reduced with version 0.7.0 of the \pywifes\ pipeline \citep{pywifes}.  \pywifes\ performs standard image pre-processing such as overscan and bias subtraction, as well as cosmic ray rejection using a version of LACosmic \citep{vandokkum01} tailored for \wifes\ data.  The wavelength solution for \wifes\ is derived using an optical model of the spectrograph which achieves an accuracy of 0.05 to 0.10\AA\ (for $R=7000$ and $R=3000$, respectively) across the entire detector.  Spectral flatfielding (i.e. correction of pixel-to-pixel quantum efficiency variations) is achieved with an internal quartz lamp, while spatial flatfielding across the full instrument field of view iss facilitated by twilight sky flats.  Once the data has been preprocessed and flat-fielded, it is resampled onto a rectilinear three-dimensional $(x, y, \lambda)$ grid (a ``data cube'').  We then flux calibrate the data cubes with the use of spectrophotometric standard stars \citep[from, e.g.,][]{oke90, bessell99, stritzinger05} and the Siding Spring Observatory extinction curve from \citet{bessell99}, while telluric features are removed using observations of smooth spectrum stars.

To further facilitate data reduction for the AWSNAP data release, we developed an SQL database for \wifes\ observations using the Python Django framework.  We created modified versions of the \pywifes\ reduction scripts that allow the user to request that all data from a specific night using a specific grating be fully reduced.  The reduction scripts then query the database for all science observations and calibration data from that night and perform the required reduction procedures.  For nights where a full suite of calibrations was not available (e.g. due to guest observations on a non-AWSNAP night), calibration solutions from the closest (in time) AWSNAP night were automatically identified via the database and employed in the reduction.  Both the wavelength solution and spatial flatfielding of the instrument are incredibly stable on long ($\sim$year) timescales (see Section~\ref{sec:wifes_performance}), thus validating the choice to use calibrations from different nights where necessary.  This new database-driven data processing mode for \pywifes\ will be an important component of the upcoming effort to develop fully robotic queue-based observing capabilities for the ANU 2.3m telescope.

The output of the \pywifes\ pipeline is a flux calibrated data cube that contains signal from the target supernova, the night sky, and occasionally the SN host galaxy.  Extraction of the SN spectrum requires isolation of the spaxels containing supernova signal and subtraction of the underlying sky (and possibly galaxy) background.  To achieve this, we constructed a custom Python-based GUI which allows the user to manually select spaxels containing the target SN (``object'' spaxels) and spaxels used to determine the background (``sky'' spaxels, which may contain some galaxy signal).  The background spectrum is determined as the median spectrum across all ``sky'' spaxels, and this median spectrum is subtracted from each ``object'' spaxel.  The final SN spectrum is then the sum of the sky-subtracted object spaxels, and the variance spectrum is the sum of the variance from all object spaxels (with no subtraction of sky variance).

This SN spectrum extraction technique produces excellent quality sky subtraction due to the robust spatial flatfielding achieved with \pywifes.  However, some obvious sky subtraction residual features are evident in the redder wavelengths of most R3000 spectra where the night sky exhibits sharp emission features from rotational transitions of atmospheric OH molecules.  The intrinsic line width of these emission lines is below the resolution of the spectrograph, meaning the observed line width is that of the spectrograph -- which in this case is only slightly larger than the detector pixel size.  The natural wavelength solution of the spectrograph shifts spaxel-to-spaxel, so when all spaxels are resampled onto the same wavelength grid this means sky lines experience pixel-wise resampling that is not uniform across the full instrument field of view.  Thus when a median sky spectrum is subtracted from a specific spaxel, some residual features arise due to this resampling effect.  A more robust technique for correcting this would be to model the intrinsic sky spectrum using the multiple samplings achieved across all spaxels and resample it to the wavelength solution of each spaxel before subtracting it \citep[as was demonstrated for two-dimensional spectroscopic data by][]{kelson03}.  Such a technique is beyond the scope of the current data release, but is being prioritised for future AWSNAP releases.

\subsection{Long-Term Behaviour of the WiFeS Instrument}
\label{sec:wifes_performance}
One key advantage of having a long-running observing program is the ability to characterise the long term behaviour of the \wifes\ instrument.  Below we analyse the stability of the wavelength solution and instrument throughput on multi-year timescales.  \wifes\ did undergo a major change in early 2013 when the detectors for both arms of the spectrograph were replaced with higher throughput E2V CCDs.  Thus we restrict our analysis to dates from late May 2013 until the end of the current data release in August 2015.


\begin{figure*}
\begin{center}
\includegraphics[width=0.95\textwidth]{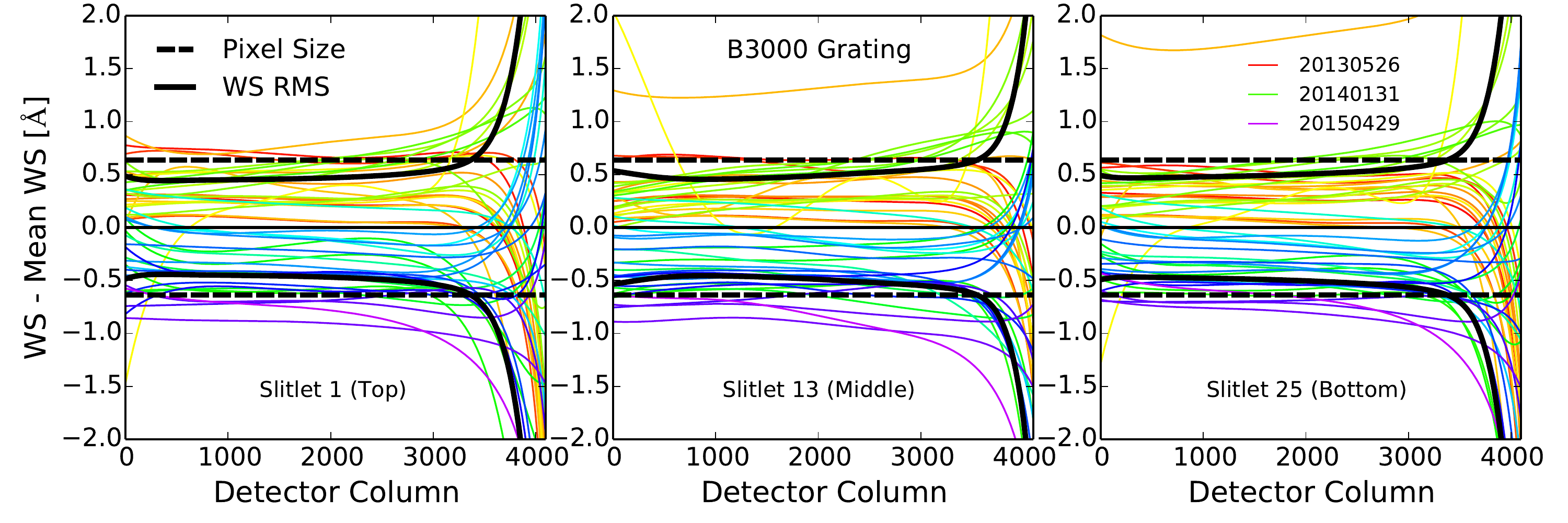}
\includegraphics[width=0.95\textwidth]{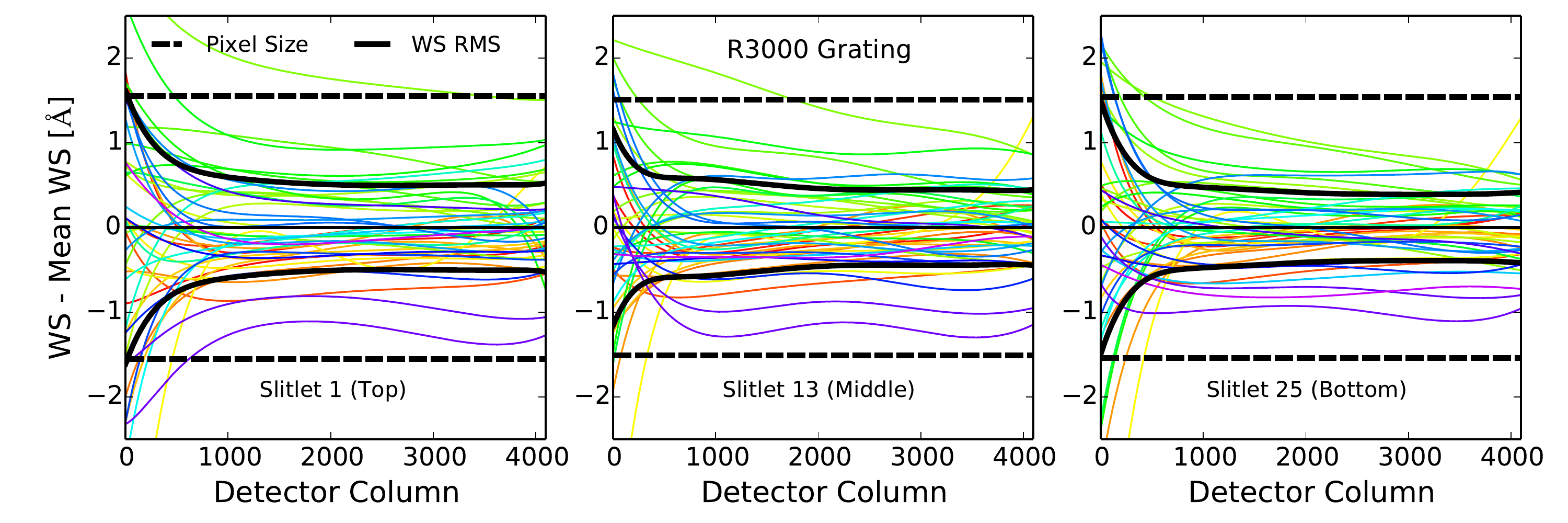}
\caption{Evolution of the \wifes\ wavelength solution over a 2-year period spanning May 2013 to April 2015.  In each panel we plot the deviation of individual wavelength solutions from the mean solution (averaged over the full 2-year period) for the top, middle, and bottom slitlets of the instrument (left, middle, and right columns, respectively) for the B3000 (top row) and R3000 (bottom row) gratings.  Wavelength solution residuals are colour-coded by date from earliest (red) to latest (purple), with the pixel size (dashed black lines) and wavelength solution residual RMS (solid black lines) displayed for comparison.}
\label{fig:wifes_wsol}
\end{center}
\end{figure*}

We collected the wavelength solution fits for the B3000 (R3000) grating from 48 (45) distinct epochs following the \wifes\ detector upgrade.  In Figure~\ref{fig:wifes_wsol} we plot the difference between the wavelength solution for each individual epoch and the mean wavelength solution across all epochs.  Epochs are colour-coded from earliest (red) to latest (purple) to illustrate potential coherent long-term shifts in the wavelength solution.  The top panels present the B3000 grating, while the bottom panels present the R3000 gratings.  The three columns represent the \wifes\ slitlets at the top, middle, and bottom of the detectors.

From these plots we see clear demonstration of the remarkable stability of the \wifes\ instrument.  The RMS variation of wavelength solution is smaller than a single pixel for both gratings (i.e. both detectors) for nearly all wavelengths, with the exception of the lower throughput regions near the dichroic boundary.  There is some evidence of a coherent shift of the blue detector wavelength solution over the two year time period probed here, but this is still less than two pixels shift.  For the red detector, we can say confidently that {\em the wavelength solution has shifted by less than a pixel over a timescale of two years}.  This remarkable stability achieved (by design) by \wifes\ means the use of the wavelength solution from an adjacent night yields negligible changes in wavelength, and thus small instrumental velocity shifts -- highly suitable for supernova analyses.


\begin{figure}
\begin{center}
\includegraphics[width=0.45\textwidth]{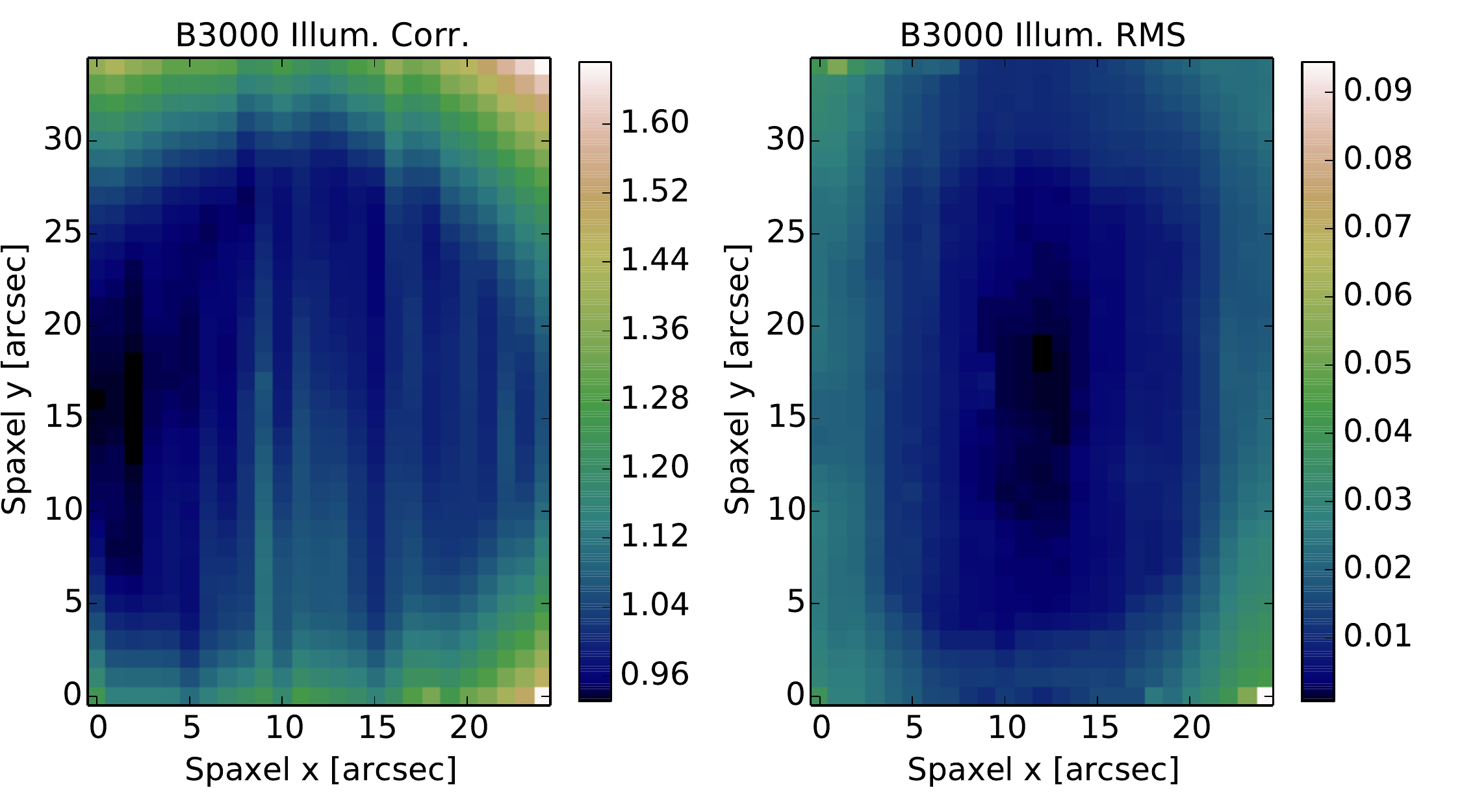}
\includegraphics[width=0.45\textwidth]{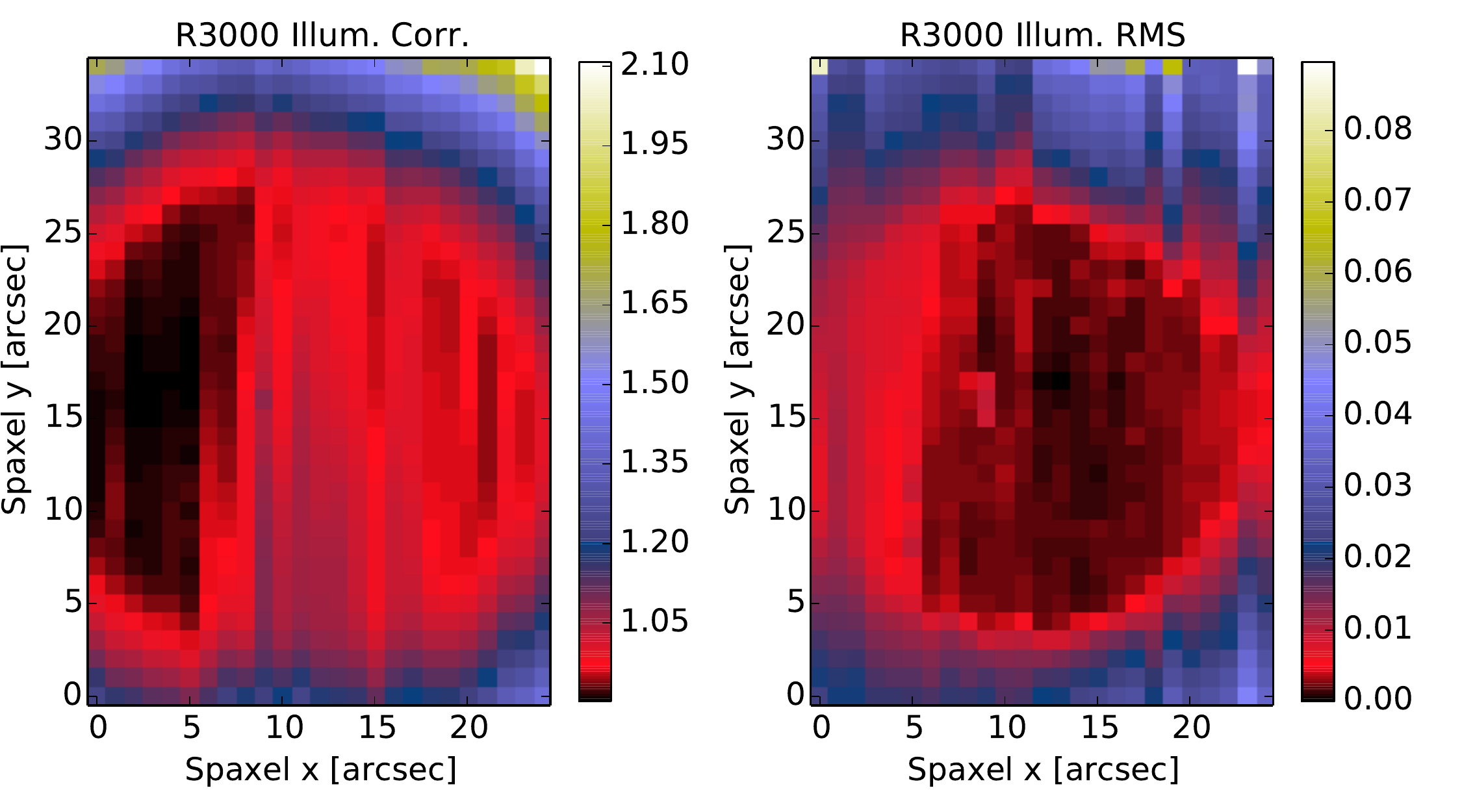}
\caption{The normalised mean (left column) and RMS (right column) of the illumination corrections for the B3000 (top row) and R3000 (bottom row) gratings.  The RMS images are shown as fractional values of the mean illumination correction.}
\label{fig:wifes_illum}
\end{center}
\end{figure}

We then collected illumination corrections (derived from twilight flat observations) from 46 (45) distinct epochs for the B3000 (R3000) grating.  In Figure~\ref{fig:wifes_illum} we show the mean illumination correction (left panels) and RMS of the illumination correction (rights panels -- presented in fractional form, i.e. the RMS divided by the mean) for both the B3000 (top) and R3000 (bottom) gratings.  We find the mean variation of the illumination correction to be 1.4\% and 1.2\% for the B3000 and R3000 gratings, respectively, for the full \wifes\ field of view.  For the innermost 8\arcsec$\times$8\arcsec\ region typically used for SN observation in AWSNAP, the RMS of the illumination correction is 0.8\% and 0.3\% for the blue and red detectors (i.e., B3000 and R3000 gratings).  Thus the throughput of the \wifes\ instrument has remarkable spatial uniformity on long (multi-year) timescales.

\subsection{Observing Conditions at Siding Spring Observatory}
\label{sec:sso_stats}
In addition to the long-term behaviour of the \wifes\ instrument, our observing program allows us to monitor the observing conditions at Siding Spring Observatory.  Below we briefly discuss the atmospheric throughput and seeing conditions experienced during AWSNAP observing.


\begin{figure}
\begin{center}
\includegraphics[width=0.45\textwidth]{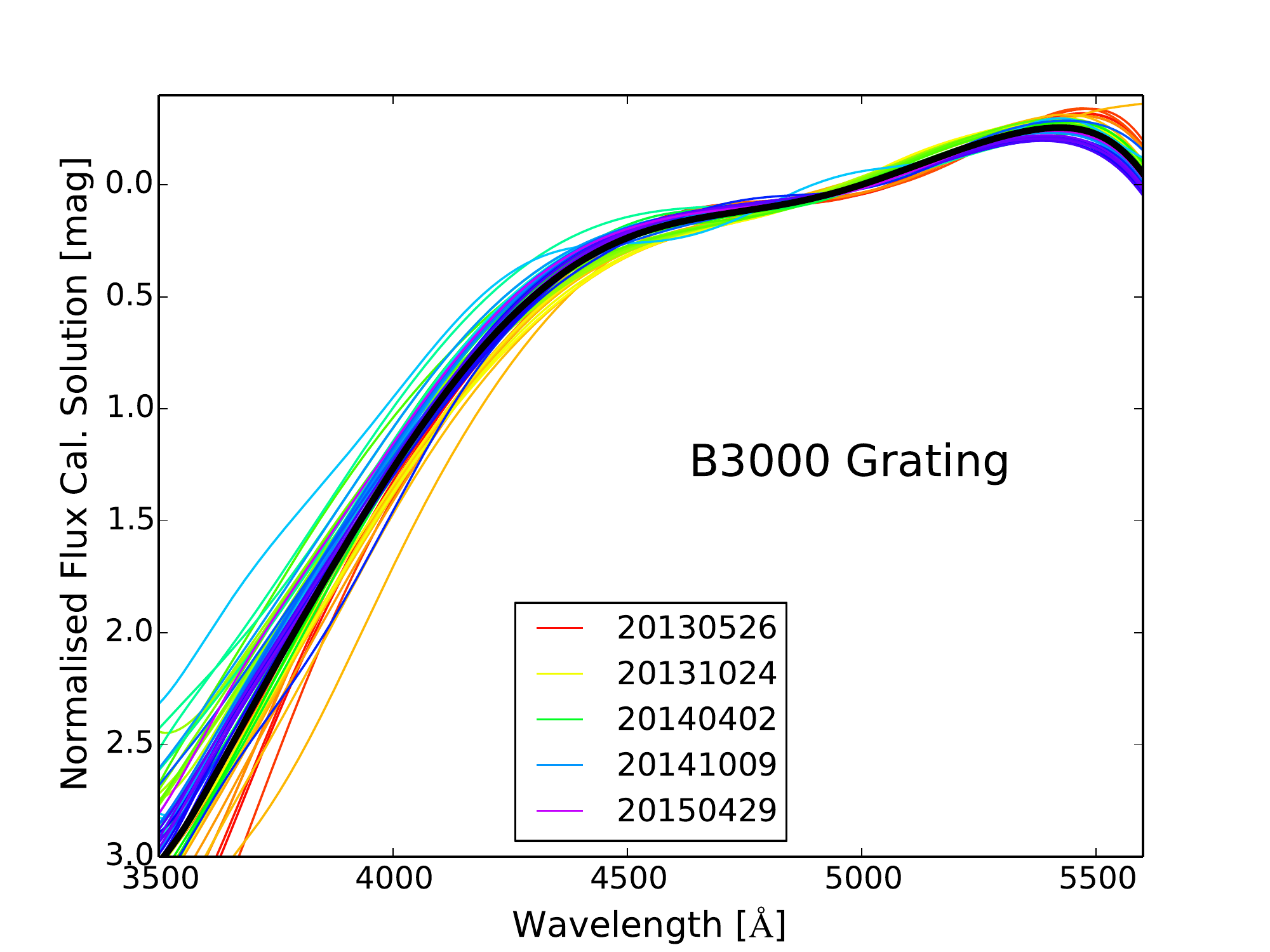}
\includegraphics[width=0.45\textwidth]{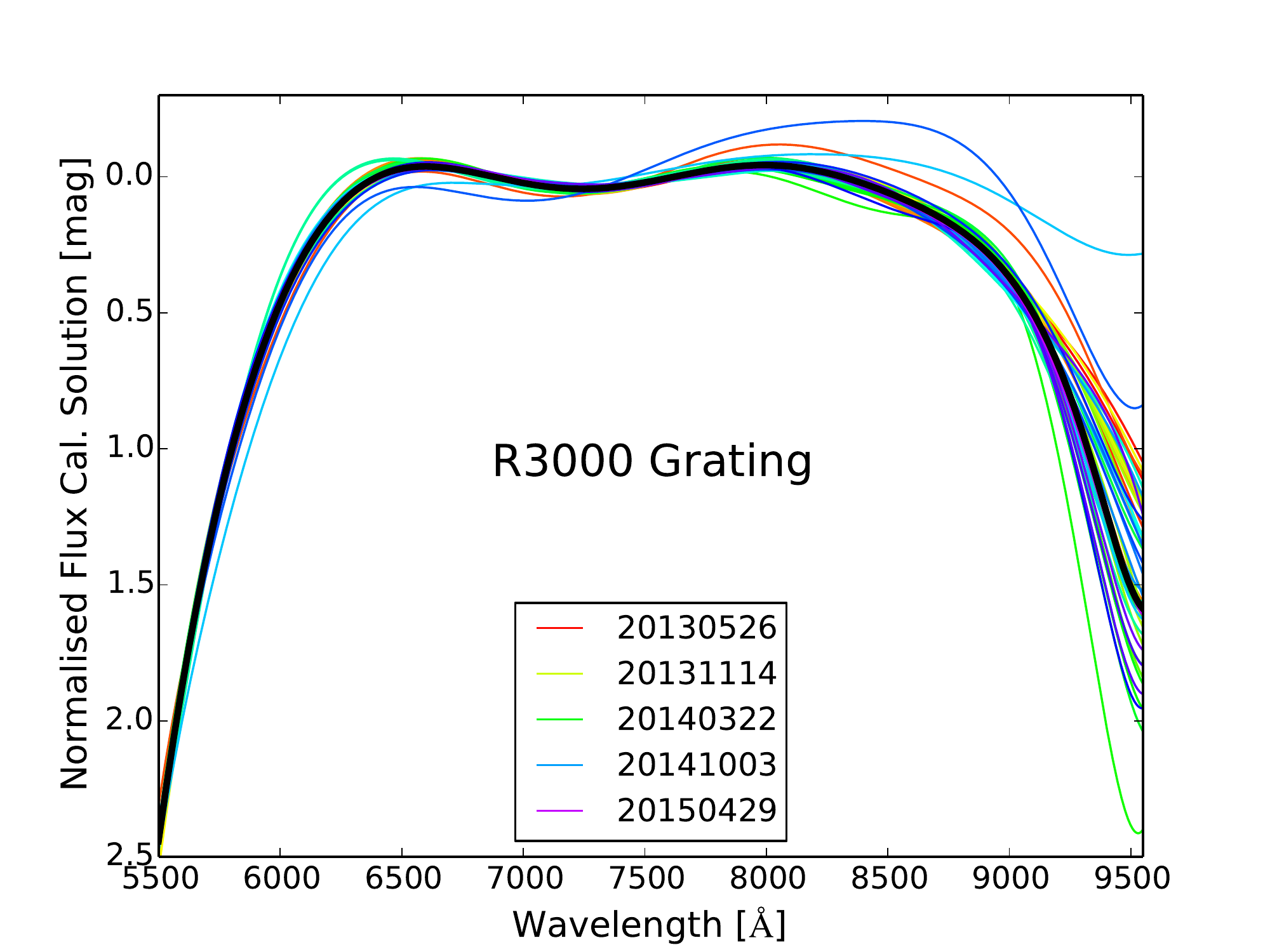}
\caption{Flux calibration solutions for the B3000 (top) and R3000 (bottom) gratings.  As in Figure~\ref{fig:wifes_wsol}, these are colour-coded by date from earliest (red) to latest (purple), with the mean flux calibration solution shown as the solid black line.}
\label{fig:wifes_flux_cal}
\end{center}
\end{figure}

We collected the flux calibration solutions from 59(54) epochs for the B3000 (R3000) grating from May 2013 to August 2015, and these are plotted in Figure~\ref{fig:wifes_flux_cal} after being normalised in the wavelength range with highest throughput (4500-5400 \AA\ and 6500-8000 \AA\ for B3000 and R3000, respectively).  These curves represent the normalised throughput (in magnitudes) of the instrument and atmosphere as measured with spectrophotometric standard stars \citep[typically from][]{oke90, bessell99, stritzinger05} whose flux has already been corrected using the nominal Siding Spring extinction curve from \citet{bessell99}.  The effects of the instrument throughput and atmospheric transmission are degenerate here, as \wifes\ instrument throughput cannot be independently measured using local (i.e. terrestrial) calibration sources.

From the curves in Figure~\ref{fig:wifes_flux_cal}, we see that the total combined throughput of the instrument plus atmosphere is relatively stable.  We calculated the RMS colour variation in the throughput curves and find variations of $\sigma(U-B) = 0.09$~mag for B3000 (note the $V$ band runs into the wavelength range where the dichroic splits light between the blue and red channels) and $\sigma(r-i) = 0.04$~mag for R3000.  We note these are calculated from the mean flux calibration solution for each night, where no attempt has been made to derive a unique extinction curve for a given night.  These colour variations are relatively small, and comparable in size to the colour dispersion found when comparing spectrophotometry to imaging photometry for other SN spectroscopy samples \citep[e.g.,][]{bsnip1, blondin12, modjaz14}.



\begin{figure}
\begin{center}
\includegraphics[width=0.45\textwidth]{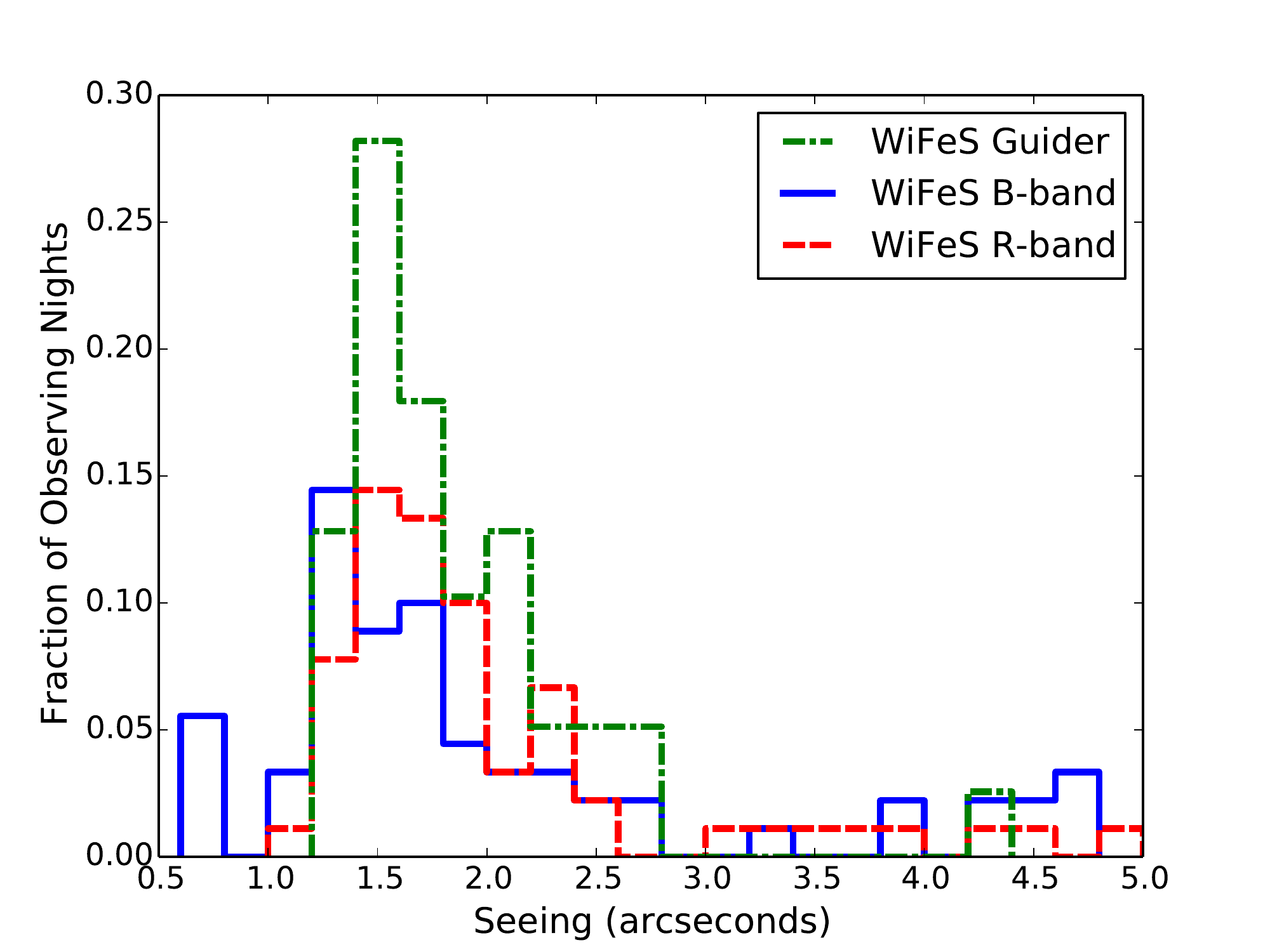}
\caption{Seeing measurements at Siding Spring Observatory as measured during AWSNAP observations.  These include measurements from the WiFeS guider camera (green dash-dot histogram), and measurements of low airmass standard star WiFeS datacubes convolved with $B$-band (blue solid histogram -- from the blue detector) and $R$-band (red dashed histogram -- from the red detector) filter curves.}
\label{fig:sso_seeing}
\end{center}
\end{figure}

We also measured the atmospheric seeing during our observing program, using both guide star FWHM measurements recorded from the \wifes\ guider camera and low airmass ($secz \leq 1.5$) \wifes\ data cubes convolved with the $B$- and $R$-band filter curves (for the blue and red cameras, respectively).  We plot the observed distribution of seeing values in Figure~\ref{fig:sso_seeing}.  The seeing distribution peaks slightly above 1.5\arcsec\ (for all measurements) with a tail predominantly filled to 2.5\arcsec, with little or no sub-arcsecond observations and a small number of observations with incredibly poor seeing (3.0\arcsec\ or greater).

\section{The AWSNAP Supernova Sample and Spectroscopic Data Release}
\label{sec:sn_sample}
In this Section we briefly describe the global characteristics of the sample of SNe comprising the first data release (DR1) for AWSNAP.  This consists of observations made between 18 July 2012 and 17 August 2015, a total of \nspec\ epochs of \nobj\ total SNe.  These spectra have all been uploaded to WISeREP \citep{wiserep}, with most made publicly available and the small remainder set to be made public with the associated PESSTO publication within the next year.  Some additional spectra taken after 17 August 2015 have been processed and released via WISeREP, and we expect the future release of AWSNAP spectra to proceed in a continuous fashion via the same procedures outlined in this work.

\begin{figure*}
\begin{center}
\includegraphics[width=0.95\textwidth]{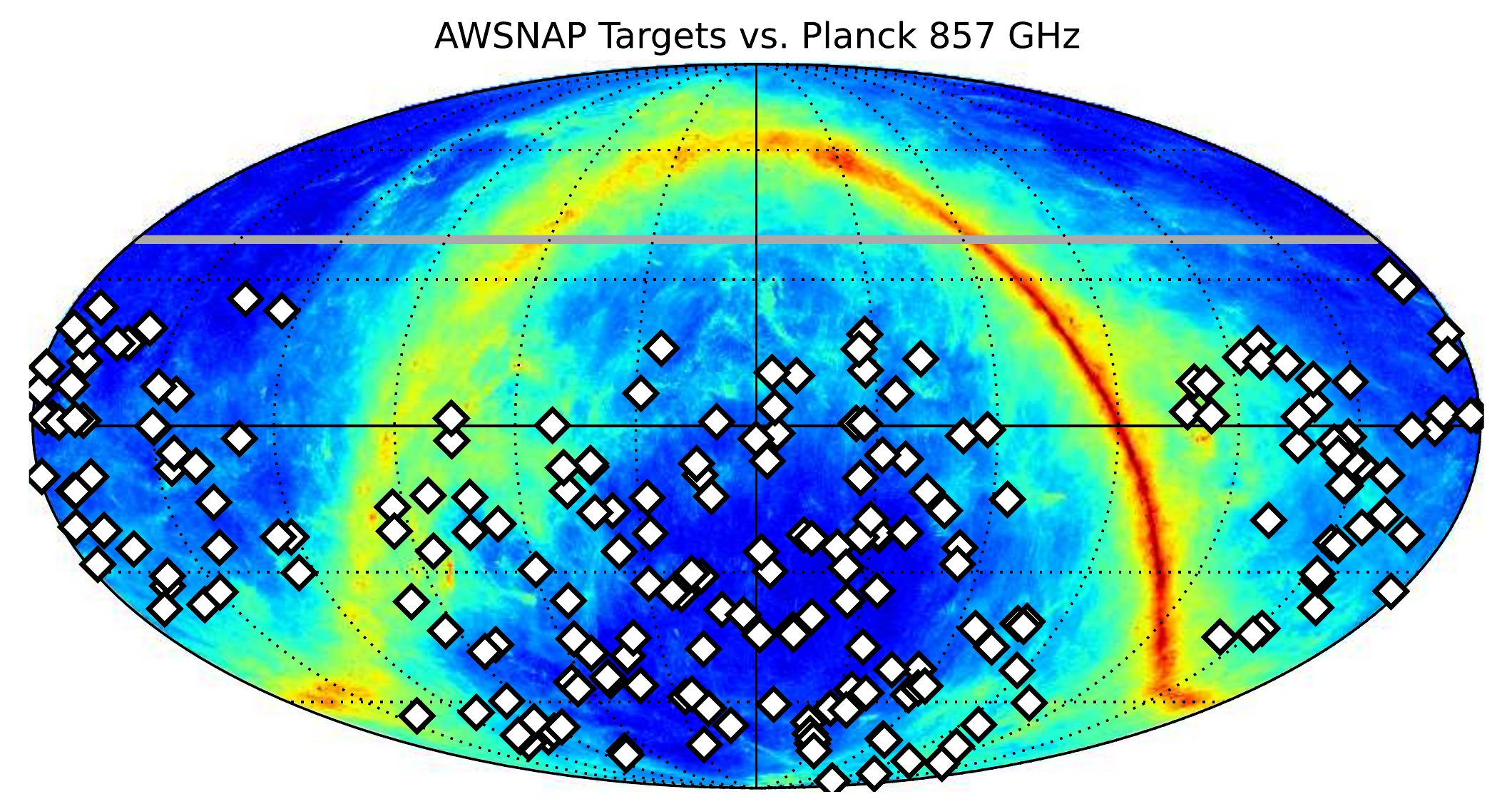}
\caption{On-sky distribution (in equatorial coordinates) of all extragalactic targets in AWSNAP DR1 (white diamonds) plotted over the 857~GHz all-sky map from the {\em Planck} satellite \citep{planck} which reveals emission by Milky Way dust.  The physical pointing limit of the ANU 2.3m telescope ($\sim+40^o$ declination) is shown as the solid grey bar.}
\label{fig:awsnap_targets}
\end{center}
\end{figure*}

The observing and target selection strategy for AWSNAP DR1 was heavily influenced by the scheduling of our observing time, which typically consisted of one single classically-scheduled full night of observing every 8-15 days throughout the entirety of the calendar year.  The most significant implication of this scheduling was that dense spectroscopic sampling (i.e., 2-3 day cadence) was generally not viable for our preferred targets.  Furthermore, additional targets were always needed to fill an entire night of observing.  Thus we frequently chose to make complementary observations of targets being observed through the PESSTO program \citep{pessto}, or chose targets whose spectroscopic data would have legacy value for future analyses.  As a result, we observed a diverse range of targets that covered the entire Southern sky.  The full list of targets and their classification information in presented in Table~\ref{tab:awsnap_targets} in Appendix~A, and their distribution on the sky is shown in Figure~\ref{fig:awsnap_targets}.

\begin{figure}
\begin{center}
\includegraphics[width=0.45\textwidth]{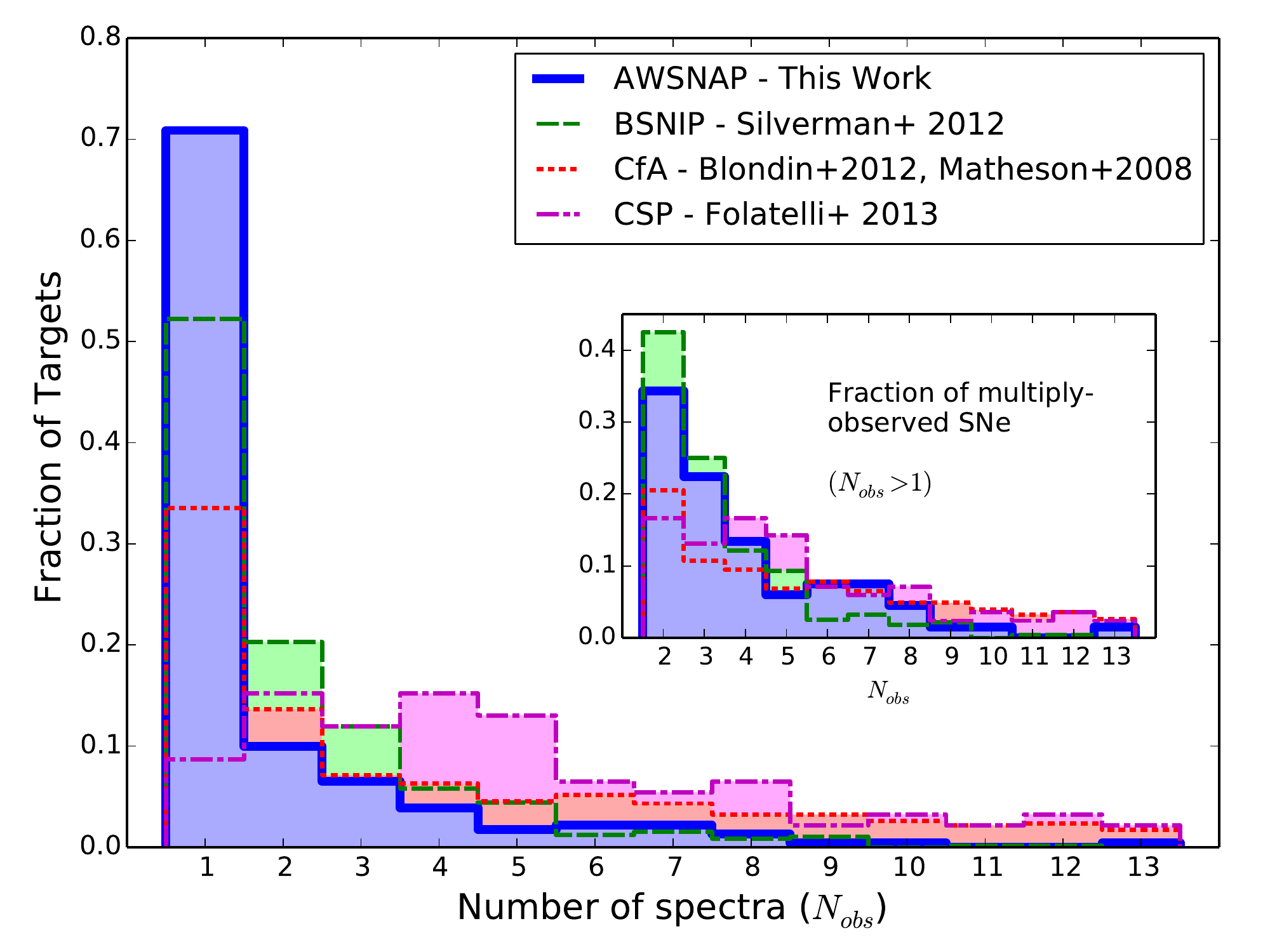}
\caption{Normalized histogram of the number of spectroscopic epochs per target for the full sample of AWSNAP DR1 targets.  For comparison we also show the same histogram for previous SN spectroscopy surveys: BSNIP \citep{bsnip1}, CfA \citep{matheson08, blondin12}, and CSP \citep{folatelli13}.  The inset shows the (re-)normalized histograms of the number of spectroscopic epochs for multiply observed targets (i.e. those with $N_{obs} > 1$) in the same surveys.}
\label{fig:nobs_hist}
\end{center}
\end{figure}

In Figure~\ref{fig:nobs_hist} we present a histogram of the number of spectroscopic observations per target for the AWSNAP sample compared to three major \snia\ spectroscopic data releases: the Berkeley SN Ia Program \citep[BSNIP;][]{bsnip1}, the Harvard Center for Astrophysics (CfA) supernova program \citep{matheson08, blondin12}, and the Carnegie Supernova Project \citep[CSP;][]{folatelli13}.  Our sample has a high fraction of singly-observed targets compared to the other programs.  This is in part due to the fact that our data release includes all our SN classication spectra, as well as the persistent need for us to fill our classically-scheduled observing queue.  We thus also show in the inset of Figure~\ref{fig:nobs_hist} the normalised histogram of spectroscopic epochs for multiply-observed SNe (i.e. targets with $N_{obs} > 2$), which shows a much more similar distribution to the other programs.

\begin{figure}
\begin{center}
\includegraphics[width=0.45\textwidth]{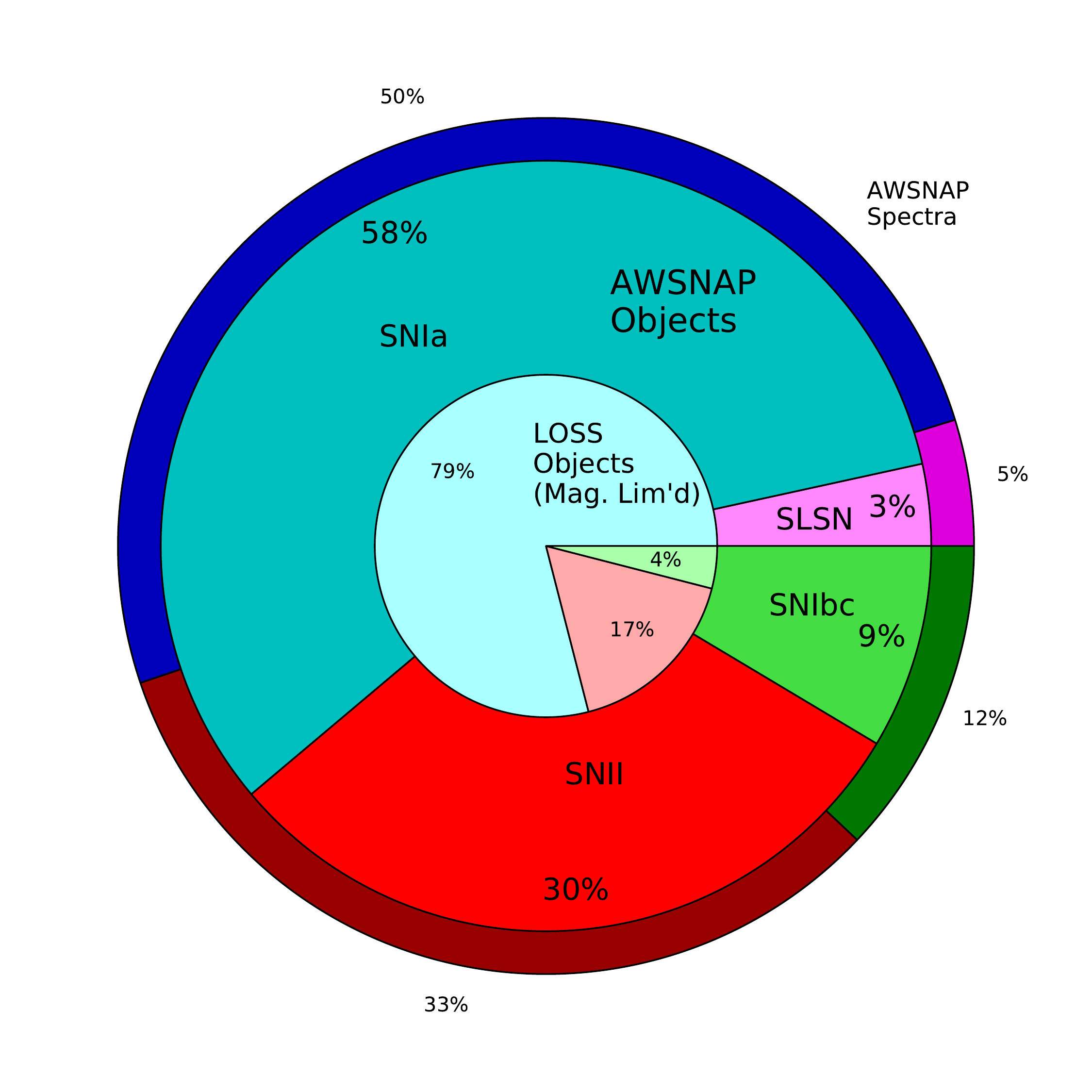}
\caption{Total number of spectra (outer ring) and SN targets (middle annulus) for AWSNAP broken down by SN type, compared to the volume-limited SN rates (inner circle) for the LOSS survey \citep[][-- note this galaxy-targeted survey did not find any SLSNe]{li11a}.  In each ring the regions are colour-coded by (broad) SN type (see text for discussion): SNe~Ia (blue), SNe~II(red), SNe Ib/Ic (green), and SLSNe (purple).}
\label{fig:sn_type_counts}
\end{center}
\end{figure}

The types of SNe observed in AWSNAP DR1 is summarized in Table~~\ref{tab:sn_type_counts} and presented graphically in Figure~\ref{fig:sn_type_counts}.  For this, we have grouped the SNe by type into four main categories:
\begin{itemize}
  \item SNe~Ia: includes the standard (``Branch-normal'') \sneia, subluminous \citep[SN~1991bg-like][]{filippenko91bg}, overluminous \citep[SN~1991T- and SN~1999aa-like][]{filippenko91T}, candidate ``super-Chandrasekhar'' \sneia\ \citep{scalzo10, scalzo12}, and SNe~Iax \citep{foley13}
  \item SNe~Ibc: SNe~Ic, SNe~Ib and SNe~IIb -- the standard classes of ``stripped-envelope'' SNe \citep[e.g.][]{bianco14, modjaz14, graur15, liu15}
  \item SNe~II: SNe~IIP and SNe~IIL -- which we note cannot be distinguished spectroscopically -- and SNe~IIn
  \item SLSNe: ``superluminous'' supernovae -- for our sample this consists entirely of the ``SLSN-Ic'' type \citep[see, e.g.,][]{inserra13} which show blue continua with weak absorption features and no hydrogen signatures
\end{itemize}
We note these groupings are made strictly to provide broad perspective on the sample statistics, but we reiterate that the SN sub-types within each group have their own unique physical mechanisms.

\begin{table}
  \caption{AWSNAP objects and spectra by SN type.}
  \label{tab:sn_type_counts}
\begin{center}
\begin{tabular}{lcc}
  \hline
SN Type & \# Objects & \# Spectra \\
\hline
SN~Ia  & 101 & 180 \\
SN~II  &  53 & 117 \\
SN~Ibc &  15 &  43 \\
SLSN   &   6 &  17 \\
\hline
{\bf Total} & 175 & 357 \\
\hline
\end{tabular}
\end{center}
\end{table}

From Figure~\ref{fig:sn_type_counts} we clearly see that, as expected, \sneia\ comprise the majority of objects and spectra in the AWSNAP sample, but comprise a smaller fraction than might be expected from a pure magnitude limited sample of SNe such as that of the Lick Observatory Supernova Search \citep[LOSS, whose relative SN rates are presented in][]{li11a}.  This may be due in part to the fact that many of the SNe in AWSNAP were discovered by untargeted supernova searches such as the All-Sky Automated Survey for Supernovae \citep[ASAS-SN;][]{asassn}.  These surveys find SNe in low mass galaxies that are missed by targeted surveys such as LOSS.  Due to the increased star-formation intensity in low mass galaxies \citep[e.g.][]{salim07} this means the relative rate of core-collapse supernovae will be higher and thus CCSNe will comprise a higher fraction of the sample.  Additionally, we explicitly targeted a higher fraction of CCSN discoveries owing to the comparative paucity of CCSN spectroscopic samples compared to \sneia\ (again arising from the magnitude-limited rates).


\section{Type Ia Supernova Spectra from AWSNAP}
\label{sec:results}
\begin{figure}
\begin{center}
\includegraphics[width=0.45\textwidth]{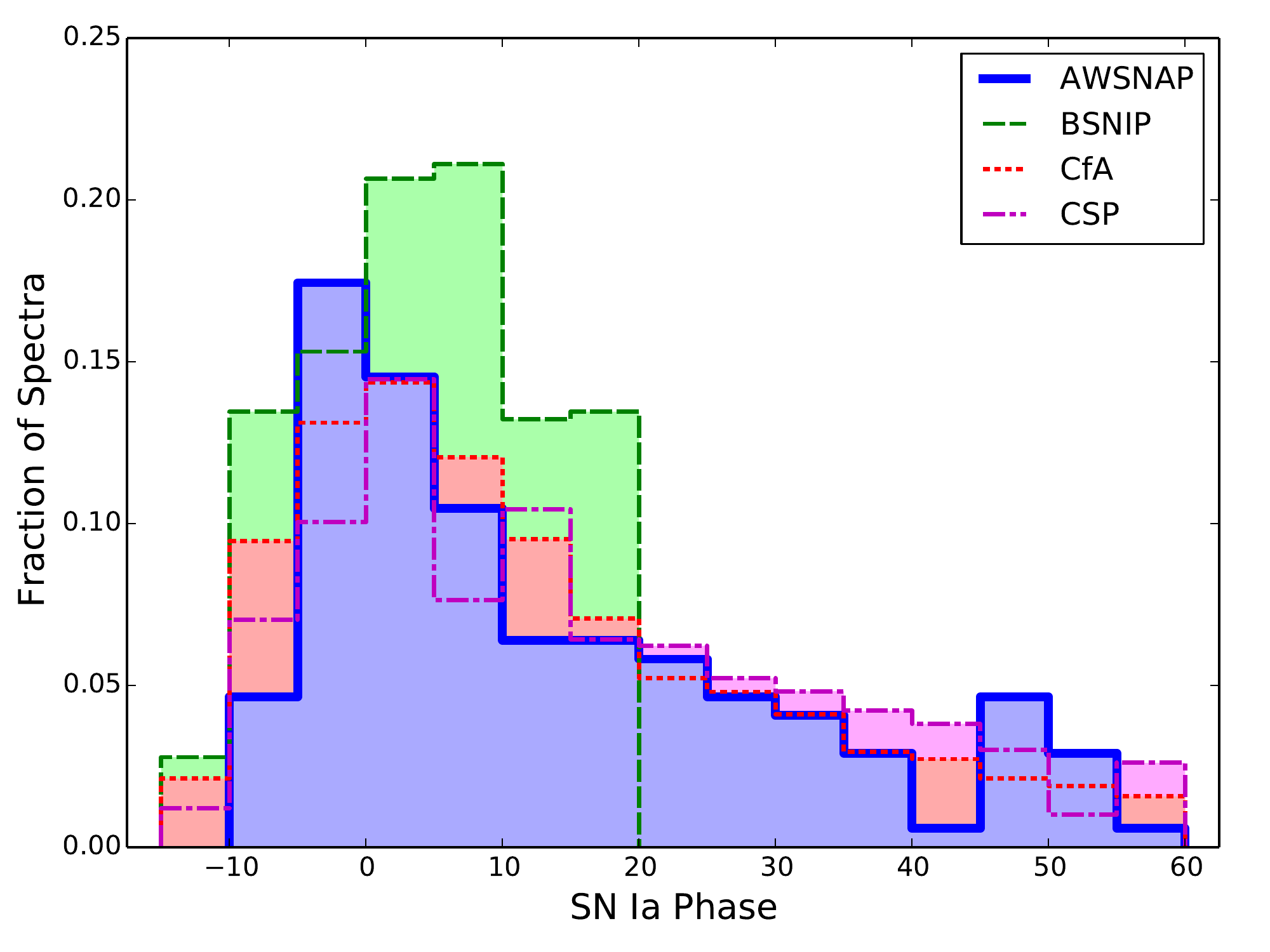}
\caption{Histograms of spectrocopic phases (with respect to $B$-band maximum light) for \snia\ spectra in AWSNAP and other \snia\ spectroscopy samples (the same as in Figure~\ref{fig:nobs_hist}).  Note the AWSNAP phases are based on the spectroscopic-based phase reported with the SN classification, which may have an associated uncertainty of 3-5 days.}
\label{fig:snia_phase_hist}
\end{center}
\end{figure}

As a first demonstration of the usefulness of the AWSNAP dataset, we analyse basic spectroscopic features of the \sneia\ in our sample -- this comprises 180 total spectra of 101 objects.  In Figure~\ref{fig:snia_phase_hist} we plot a histogram of the phases\footnote{Based on the spectroscopically-estimated phase reported at the time of classification.} of our \snia\ spectra, again compared to previous large \snia\ spectroscopy surveys (CfA, BSNIP, CSP).  Our sample shows a similar phase coverage as previous surveys, with perhaps a slight increase in earlier phases due to improved SN discovery efficiency from nearby SN searches.  In the sections that follow we further analyse features of interest from the AWSNAP \snia\ spectroscopy sample.

\subsection{Spectral features in \sneia\ at maximum light}
\label{sec:snia_bmax}
We begin by inspecting the spectra of \sneia\ close to the epoch of peak brightness.  These maximum light spectra have a long history of providing key insights into the diversity of \snia\ explosions through the study of ``spectral indicators'' \citep{nugent95, hatano00, benetti05, bongard06, bongard08, hachinger06, hachinger08, branch06, branch09, bronder08, bsnip2}, as well as potential avenues for improving the cosmological standardization of \sneia\ \citep{wang09, bailey09, blondin11, bsnip3}.  Thus we actively targeted many \sneia\ (which had already been classified) on an epoch close to maximum light to obtain high quality spectra facilitating such studies.

We isolated the sample \sneia\ with a spectrum within 5 days of estimated peak brightness.  It is important to reiterate here that the phases for our \snia\ sample are extrapolated from the reported spectroscopic classification phase and date.  It has been demonstrated previously \citep[e.g.][]{blondin12} that the phase determined for \sneia\ via spectroscopic matching codes such as SNID \citep{snid} typically has an uncertainty of at least 3-5 days.  Thus we may have included \snia\ spectra as much as 10 days removed from maximum light.  Our analysis here is intended to be illustrative of the utility of our published spectra, and more detailed quantitative spectroscopic analyses should always be coupled to a robust photometric data set.

For the analysis that follows, we measure two key quantities of interest: the silicon absorption ratio $R_{Si}$, and the strength of high-velocity features (HVFs)  $R_{HVF}$.  For this work, we define $R_{Si}$ as the ratio of the pseudo equivalent width (pEW) of the 5962\AA\ feature to the pEW of the 6355\AA\ feature.  The pEW for each feature is measured by fitting a linear pseudo-continuum across narrow regions redward and blueward of the given feature \citep[similar to methods employed in][]{bsnip2}, then integrating the flux in the normalized absorption feature.

$R_{Si}$ is known to correlate strongly with the \snia\ light curve decline rate \citep{nugent95}.  We quantified the relationship between our definition of $R_{Si}$ (the pEW ratio) and light curve decline rate $\Delta m_{15}$ by fitting a linear relationship between these quanities for a sample of 342 \sneia\ collated from the CfA, BSNIP, and CSP samples.  From these data we derive:
\begin{equation}
  \Delta m_{15} = 2.04 R_{Si} + 0.808 \pm 0.190
\end{equation}
Below we will use this relation to display the corresponding range of $\Delta m_{15}$ spanned by our observed values of $R_{Si}$.

We measure the HVF strength for our \snia\ sample using the same techniques employed in \citet{childress14}, and similarly define $R_{HVF}$ as the ratio of the pEW of the high-velocity feature (HVF) to the pEW of the photospheric velocity feature (PVF).  Briefly, we first normalise the Ca NIR feature using a linear pseudo-continuum fit.  We then fit the Ca NIR feature with two Gaussians in {\em velocity space} with velocity centres, widths, and absorption depths fitted with a Python {\tt mpfit} routine.  Each component in velocity space corresonds to a triplet in wavelength space, and we set the absorption depth of each component equal to each other (i.e. the optically thick limit).  As in \citet{childress14}, we require the photospheric velocity component to have a velocity centre within 20\% of the value derived for the Si 6355\AA\ feature, though in most cases the results are the same if this requirement is removed.


\begin{figure}
\begin{center}
\includegraphics[width=0.45\textwidth]{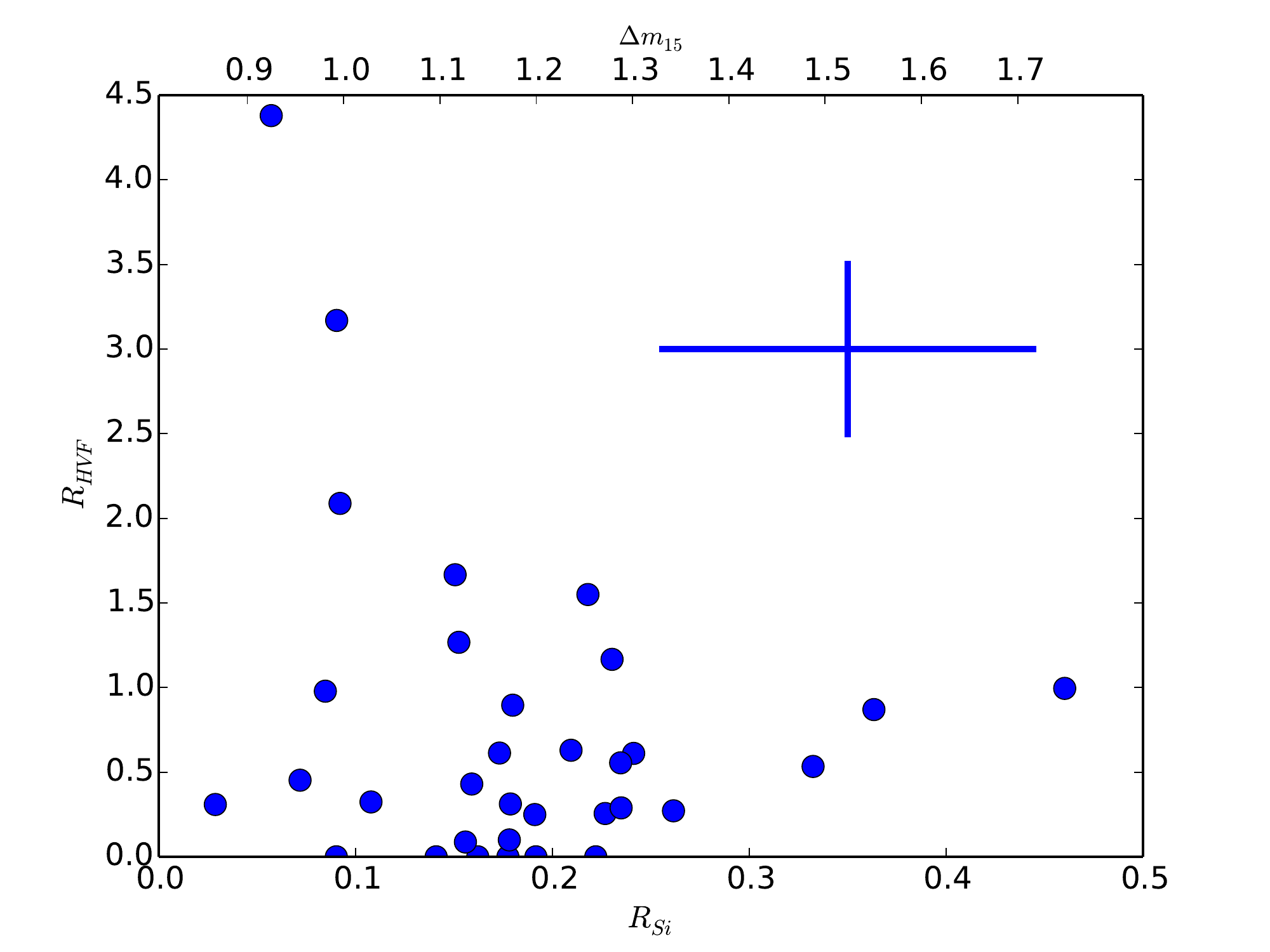}
\caption{Strength of the high-velocity features (HVFs) in the Ca~II NIR triplet -- using the quantity $R_{HVF}$ as defined by \citet{childress14} -- plotted against the Si~II absorption strength ratio $R_{Si}$ defined by \citet{nugent95}.  On the top axis we show the rough equivalent light curve decline rate  $\Delta m_{15}$ values corresponding to the range of $R_{Si}$ values.  The crosshair in the upper right represents the characteristic errors in measurement of $R_{HVF}$ (typically 20\%, plotted here for the larger values of $R_{HVF}$) and the error in  $\Delta m_{15}$ when converted from $R_{Si}$.}
\label{fig:hvfs_snia}
\end{center}
\end{figure}

In Figure~\ref{fig:hvfs_snia} we plot the measured high velocity feature strength ($R_{HVF}$) versus the silicon absorption strength ratio $R_{Si}$.  For illustrative purposes, we also show the scale of light curve decline rate $\Delta m_{15}$ corresponding to the plotted range of $R_{Si}$ using the above relation.  A representative errorbar for $\Delta m_{15}$ from the above relation and a typical $R_{HVF}$ errorbar are shown in the figure.

Figure~\ref{fig:hvfs_snia} demonstrates a tendency for \sneia\ with strong HVFs to also have low values of $R_{Si}$ and thus broad light curves (low $\Delta m_{15}$).  This correlation of HVFs with \snia\ light curve width was first observed by \citet{maguire12} in composite high-redshift \snia\ spectra, and subsequently confirmed by numerous studies of low-redshift \sneia\ \citep{childress14, maguire14, pan15, silverman15, zhao15}.  Our results support these studies with spectra alone.

\subsection{Narrow sodium absorption features in \snia\ spectra}
\label{sec:sodium}
A great advantage of the higher resolution of \wifes\ (compared to many spectrographs deployed for other SN spectroscopy surveys) is the ability to detect narrow absorption features, particularly the Na~I doublet at $\lambda\lambda5890/5896$\AA.  This feature has a long history of being used to infer the presence of foreground dust in \sneia\ \citep{barbon90, turatto03, poznanski11, poznanski12, phillips13}.  Earlier works attempted to derive correlations between \snia\ colors and sodium absorption strength (i.e. the absorption equivalent width), but \citet{poznanski11} showed sodium to be a poor indicator of \snia\ reddening.  \citet{phillips13} refined this result by showing that \snia\ reddening exhibits a strong correlation with absorption strength of the diffuse interstellar bands (found exclusively in the interstellar medium) but some \sneia\ have an excess of sodium absorption that does {\em not} coincide with associated reddening of the SN.  Thus sodium features in \sneia\ remain instructive but should be considered with a measure of caution.

Recently, velocities of sodium absorption features in \snia\ spectra have been used as a diagnostic of circumstellar material (CSM), including cases of some \sneia\ \citep{patat07, simon09} where the sodium absorption features exhibit variability \citep[though most do not -- see][]{sternberg14}, indicating SN-CSM interaction.  More recently, a statistical analysis of the velocity distribution of sodium features in \sneia\ by \citet{sternberg11} found an excess of \sneia\ with blueshifted sodium features, indicating that a fraction of \sneia\ explode inside an expanding shell of material that was presumably shed by the \snia\ progenitor system prior to explosion.   \citet{maguire13} extended this work to show that the excess of \sneia\ with blueshifted sodium absorption was populated predominantly by the more luminous \sneia\ with slow declining light curves (i.e. high ``stretch'').

We thus searched for the presence of sodium absorption features in our sample of \snia\ spectra.  For most \sneia\ (13) this was done with the lower resolution (R3000) observations of the SN at maximum light.  For six \sneia, we also had a higher resolution (R7000) observation of the SN at maximum light.  In a few instances this observation was triggered by the presence of strong reddening being reported in the SN classification announcement.  For the other instances, we obtained both a low-resolution (R3000) and high-resolution (R7000) spectrum in the red while obtaining a longer exposure blue (B3000) spectrum during nights near full moon when the blue sky background was exceptionally high.

\begin{figure}
\begin{center}
\includegraphics[width=0.48\textwidth]{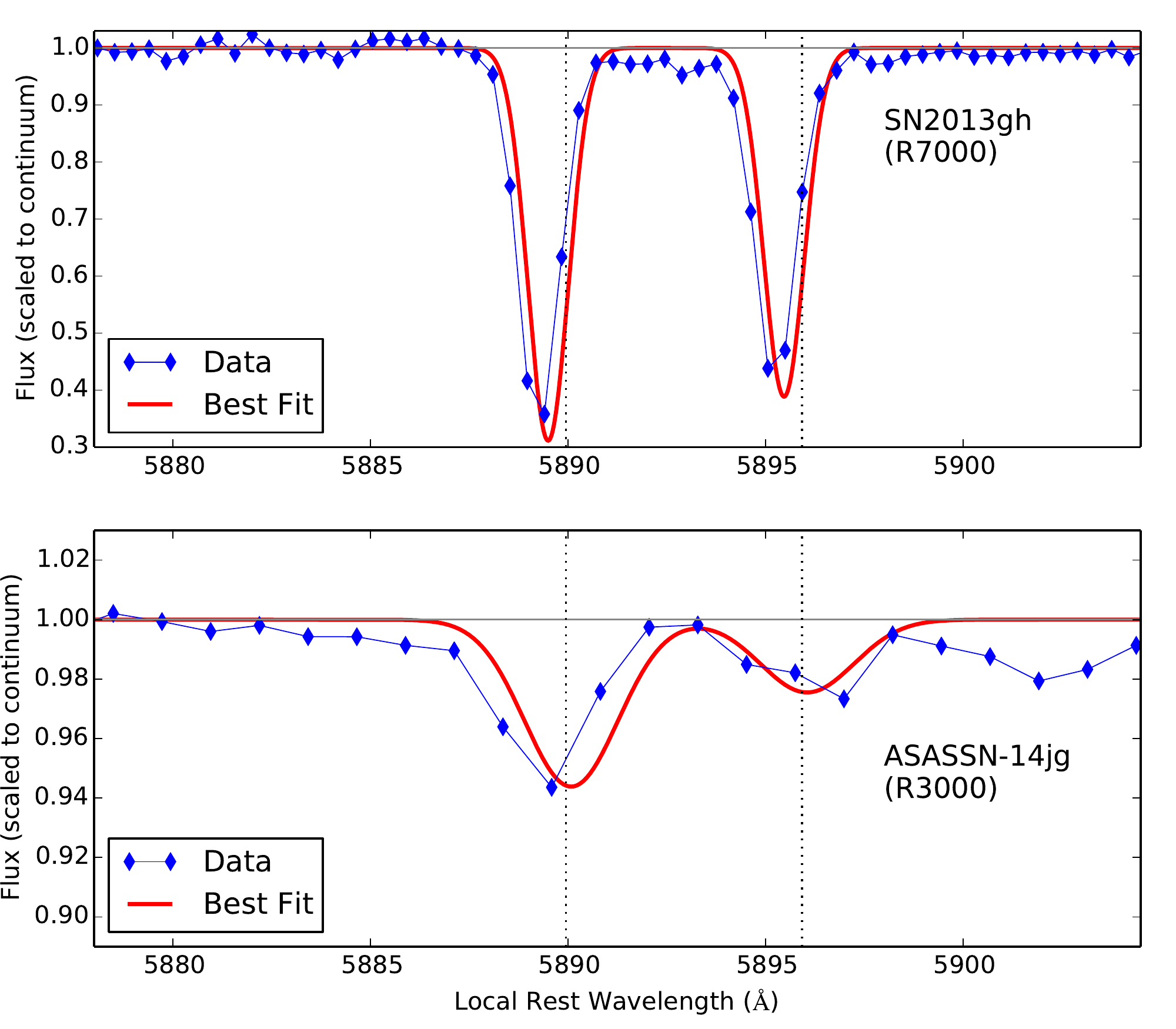}
\caption{Sodium absorption fit examples for both the R7000 (top) and R3000 (bottom) gratings.  Data (which have been normalised to the local continuum fit) are shown as blue diamonds while the best fit absorption profile is shown as the solid red curve.  For reference, we also mark the continuum level (horizontal black line at value 1.00) and the rest wavelengths of the sodium doublet (vertical dotted grey lines).}
\label{fig:nad_fits}
\end{center}
\end{figure}

We fitted the sodium doublet absorption profile as follows.  First the spectrum was normalized to the local continuum by fitting a quadratic to the SN flux between 10-25\AA\ redward or blueward of the doublet centre (5893\AA).  The absorption profile was fitted as two Gaussians with rest wavelengths set by the doublet wavelengths but with unknown (common) velocity shift and velocity width and (independent) absorption depths.  This was done with an {\tt mpfit} routine in Python, which accounts for variable covariances and returns the appropriate fit values and uncertainties.  We show representative examples for fits to data from both gratings in Figure~\ref{fig:nad_fits}.  The outcomes for our sodium profile fits are presented in Table~\ref{tab:nad_fit_results} in Appendix~A, and comprise 6 successful fits for targets with R7000 spectra, and 13 for R3000 spectra. 

The primary quantity we wanted to measure from the sodium absorption feature was its velocity with respect to the local standard of rest at the SN site.  The default local rest velocity was initially set to be the systemic velocity of the SN host galaxy.  In some cases we detected clear nebular emission lines at the SN site present in the SN spectrum -- for these cases the local rest velocity (i.e., the rotational velocity of the host galaxy at the site of the SN) was measured from the H$\alpha$ emission line.

In two cases (SN~2014ao and ASASSN-14jg) no local H$\alpha$ emission was present in the SN spectrum, and the SN sodium velocity differed from the host systemic velocity by more than 100 \kms.  To obtain the true local rest velocities for these two \sneia, we took advantage of the integral-field data provided by \wifes, which allows us to measure host galaxy properties (such as velocity) over a broad field of view.

\begin{figure*}
\begin{center}
\includegraphics[width=0.95\textwidth]{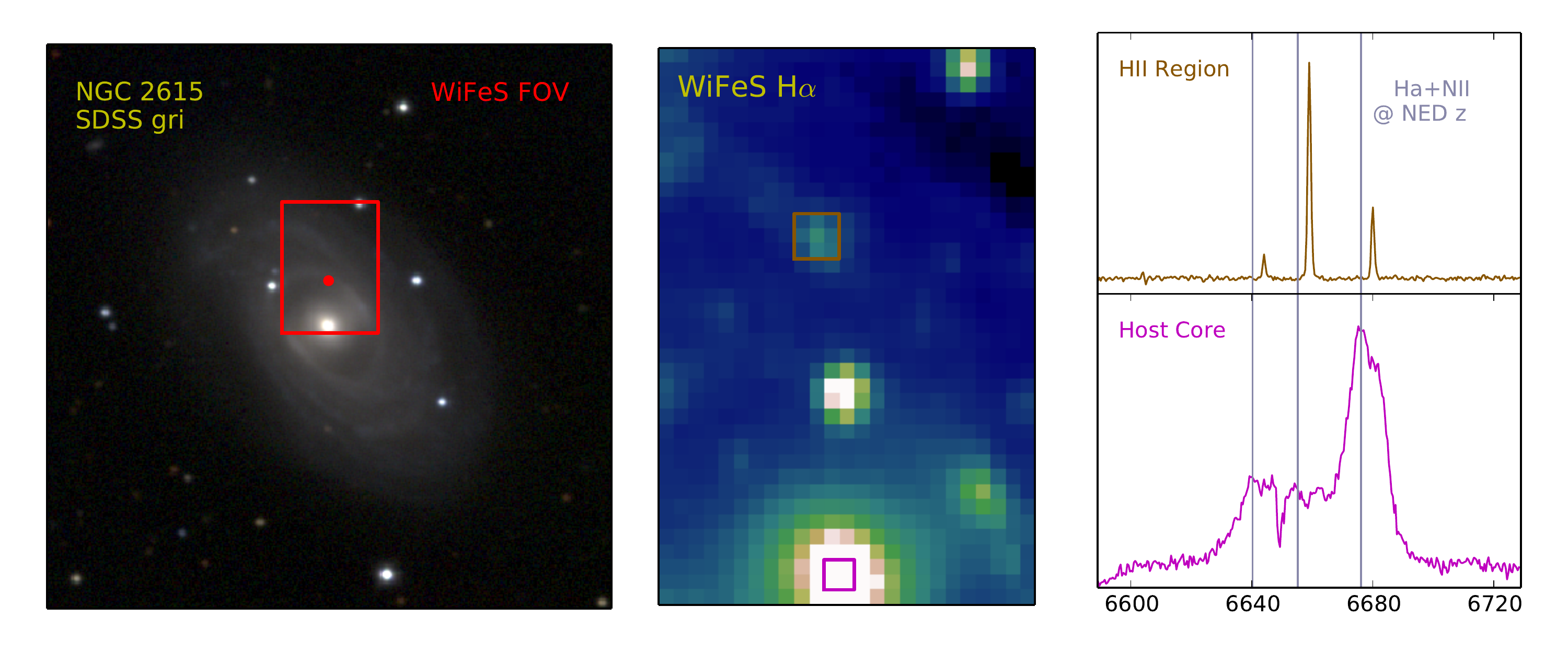}
\caption{Determination of the local velocity for SN~2014ao in NGC~2615 with \wifes.  Left: SDSS \citep{york00} {\em gri} color composite -- created with SWARP \citep{swarp} and STIFF \citep{stiff} -- with the \wifes\ field of view (red rectangle) and SN location (red dot) highlighted.  Middle: Image of the SN~2014ao \wifes\ data cube in the isolated wavelength range within $\pm6$~\AA\ (i.e. $\pm300$~\kms) of the wavelength of H$\alpha$ at the published redshift of NGC~2615, with host core (purple square) and nearby H\,\textsc{ii} region (brown square) highlighted -- the SN is the bright object near the centre.  Right: Extracted \wifes\ spectra of the nearby H\,\textsc{ii} region (top) and host core (bottom) near the H$\alpha$+NII emission line group, with the expected location of those lines at the published redshift of NGC~2615 \citep[z=0.014083][]{zNGC2615} shown as the vertical grey lines.}
\label{fig:sn2014ao}
\end{center}
\end{figure*}

We illustrate this for SN~2014ao in Figure~\ref{fig:sn2014ao}: the \wifes\ field of view extends from the host galaxy core to the outer edge of its spiral arms in the direction of the SN.  We were able to extract the velocity of an H\,\textsc{ii} region along the SN-host axis and found its velocity differed significantly from that of the host core.  As galaxy velocity curves tend to flatten at large radii, we use the H\,\textsc{ii} region velocity as the local rest velocity of SN~2014ao -- a value much closer to the measured sodium absorption velocity.


\begin{figure}
\begin{center}
\includegraphics[width=0.45\textwidth]{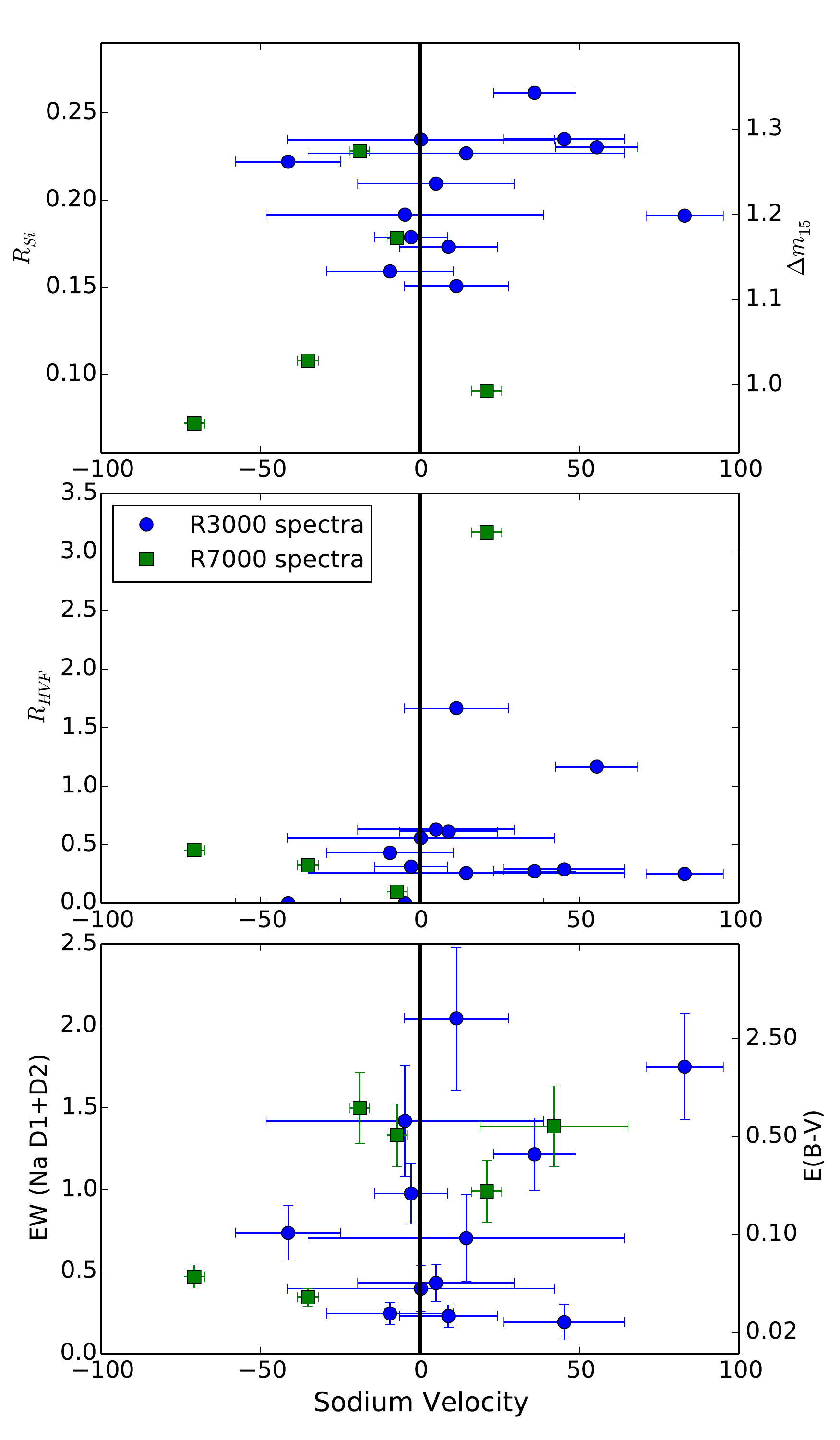}
\caption{Top: Silicon absorption ratio $R_{Si}$ plotted against velocity centre of the narrow sodium absorption feature (as in Figure~\ref{fig:hvfs_snia} we show the corresponding values of $\Delta m_{15}$, though note the smaller range). Middle: HVF strength ($R_{HVF}$) plotted against sodium absorption velocity.  Bottom: Absorption equivalent width of the combined D1+D2 sodium lines plotted against sodium absorption velocity. On the right axis of this panel we use the relation of \citet{poznanski12} to show the reddening values $E(B-V)$ corresponding to the measured sodium equivalent widths if the absorption arises solely from the ISM -- though this is unlikely to be true for all \sneia\ \citep[][-- see discussion in text]{poznanski11, phillips13}. In all panels, higher resolution observations with the R7000 grating are displayed as green squares, while lower resolution R3000 observations are shown as blue circles.}
\label{fig:sodium_results}
\end{center}
\end{figure}

With the final sodium absorption velocities for our sample, we can inspect the relationship between sodium velocity and other spectroscopic properties of our \snia\ sample.  In Figure~\ref{fig:sodium_results} we plot the silicon absorption ratio ($R_{Si}$), the HVF absorption strength ($R_{HVF}$), and the absorption strength (i.e. equivalent width) of sodium itself against the velocity of the sodium absorption feature.  Based on results of \citet{maguire13}, we would expect \sneia\ with low $R_{Si}$ and high $R_{HVF}$ values (the high stretch, slow declining \sneia) to have a slight excess of blueshifted sodium absorption.  Our sample size here is too small (and not well-selected) to make any robust statement about such preferences.  However we note that this analysis was not the explicit objective of our observations, but instead was a supplemental outcome facilititated by the nature of the \wifes\ data.


Thus \wifes\ is an excellent instrument for measuring sodium absorption in \sneia, a key observable for investigating \snia\ progenitor systems.  This is particularly true for observations taken with the R7000 grating, which provides sodium velocity uncertainties of order a few \kms, thus enabling a robust classification of the SN as being ``blueshifted'' or ``redshifted''.  More importantly, perhaps, is the capability of \wifes\ to observe a wide field of view around the SN.  This enables a measurement of the local systemic velocity at the SN location, even in cases where emission from the host galaxy is weak at the SN location itself.

\subsection{Spectroscopic evolution of SN~2012dn}
\label{sec:sn2012dn}
\begin{figure}
\begin{center}
\includegraphics[width=0.48\textwidth]{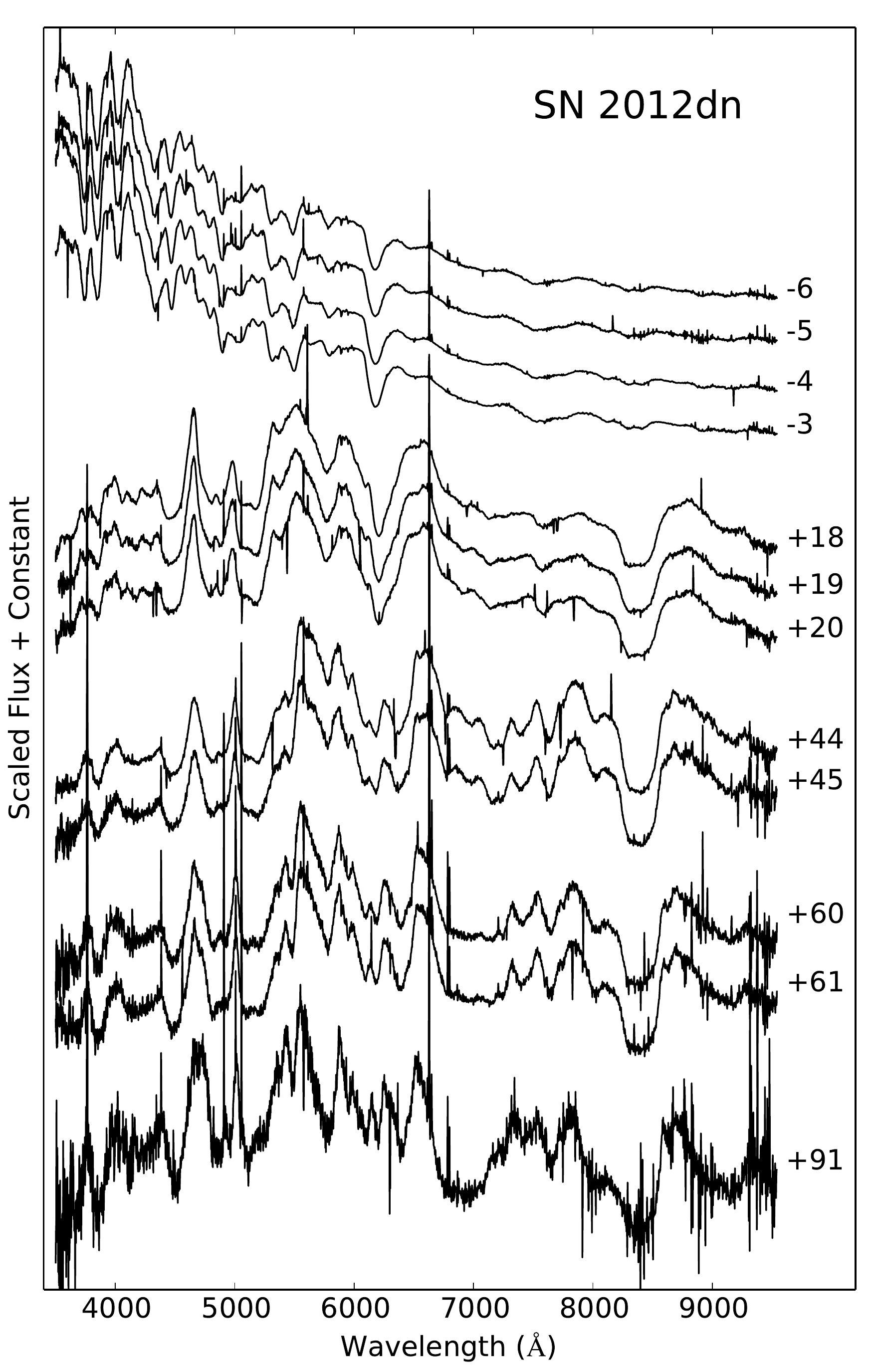}
\caption{AWSNAP time series of SN~2012dn, labeled by phase with respect to the date of maximum light \citep[2012 July 24, as determined by][]{chakradhari14}.  Note these observations come from the first semester of AWSNAP when observing time was allocated in multi-night blocks separated sometimes by a month or more.}
\label{fig:sn2012dn_time_series}
\end{center}
\end{figure}

The SN with the greatest number of spectroscopic epochs in AWSNAP is SN~2012dn, and we show its AWSNAP spectroscopic time series in Figure~\ref{fig:sn2012dn_time_series}.  SN~2012dn was a spectroscopically peculiar \snia\ whose photometric and spectroscopic evolution was studied extensively by and \citet{chakradhari14} -- they found it to be similar to the candidate ``super-Chandrasekhar'' (hereafter super-Chandra) \snia\ SN~2006gz \citep{hicken07}.  We comment on only a few additional outcomes from the AWSNAP data, but refer readers to \citet{chakradhari14} and \citet{parrent16} for a thorough discussion of this interesting object.

\begin{figure}
\begin{center}
\includegraphics[width=0.48\textwidth]{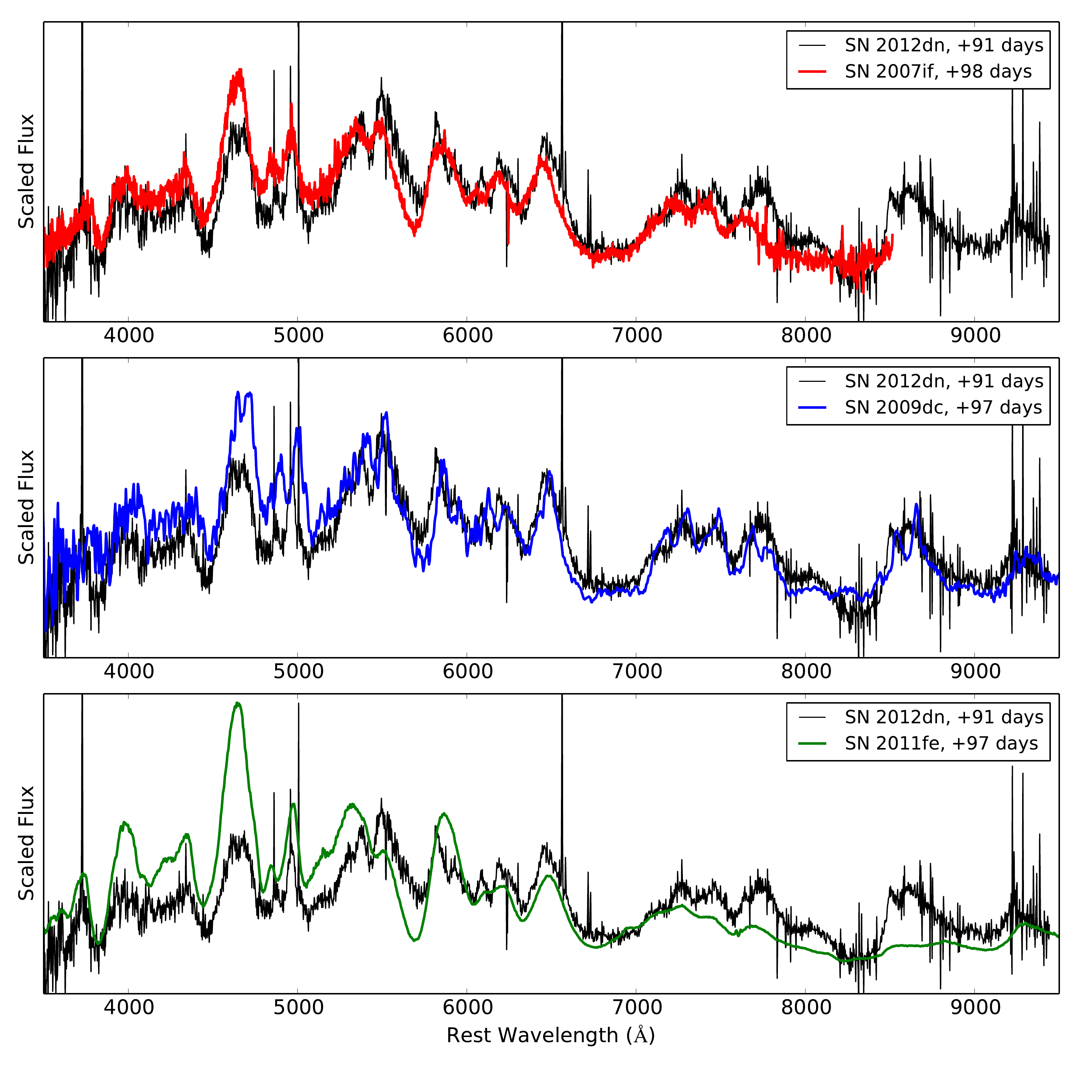}
\caption{SN~2012dn at its latest AWSNAP epoch (+91 days on 2012 Oct 23) compared to other candidate super-Chandra \sneia\ SN~2007if at +98 days \citep[top panel, from][]{silverman11} and SN~2009dc at +97 days \citep[middle panel, from][]{taubenberger11}, as well as the normal SN~2011fe \citep[bottom panel, from][]{pereira13}.}
\label{fig:sn2012dn_late}
\end{center}
\end{figure}

We obtained a very high signal-to-noise spectrum of SN~2012dn at phase +91 days \citep[with respect to the date of maximum light 2012 July 24, as determined by][]{chakradhari14}.  At this epoch the SN is beginning to enter the nebular phase when the ejecta become optically thin, revealing emission from the iron group elements (IGEs) near the centre of the SN.  Spectra at these epochs provide an excellent diagnostic of the nucleosynthetic products of the SN explosion.  In Figure~\ref{fig:sn2012dn_late} we present our +91 day spectrum of SN~2012dn compared to very late spectra of other candidate super-Chandra \sneia\ SN~2007if \citep{scalzo10, yuan10, taubenberger13} and SN~2009dc \citep{silverman11, taubenberger11, yamanaka09, tanaka10, hachinger12, kamiya12, taubenberger13}, as well as the gold standard normal \snia\ SN~2011fe \citep{nugent11, li11fe, parrent12, pereira13}.

This comparison clearly reveals a strong spectroscopic similarity between SN~2012dn and the candidate super-Chandarsekhar \sneia, and a distinct dissimilarity with SN~2011fe.  Perhaps most prominent is the weaker Fe~III line complex at $\sim4700$~\AA\ for the super-Chandra \sneia\ compared to SN~2011fe.  This discrepancy is also evident in fully nebular spectra of super-Chandra \sneia\ at $\sim1$~year past maximum light, as discussed by \citet{taubenberger13}.  This indicates that the ionisation state of the super-Chandra \sneia\ at these phases is different from normal \sneia\ -- whether the diminished Fe\,\textsc{iii} emission arises from a higher or lower average ionisation state remains uncertain.

Additionally, the velocity profile of the emission features in the super-Chandra \sneia\ exhibits a marked difference to that of SN~2011fe (and other normal \sneia).  The normal \snia\ profile appears very Gaussian \citep[and indeed is generally well fit by a Gaussian profile --][]{childress15}, while the super-Chandra velocity profile appears sharper at the centre.  This is particularly evident for the line features at $\sim$5900~\AA\ and $\sim$6300~\AA, which at these epochs are dominated by Co\,\textsc{iii}.  Further spectral modeling of this and other late-phase super-Chandra spectra may reveal important insights into the structure and composition of the super-Chandra ejecta.

\begin{figure}
\begin{center}
\includegraphics[width=0.48\textwidth]{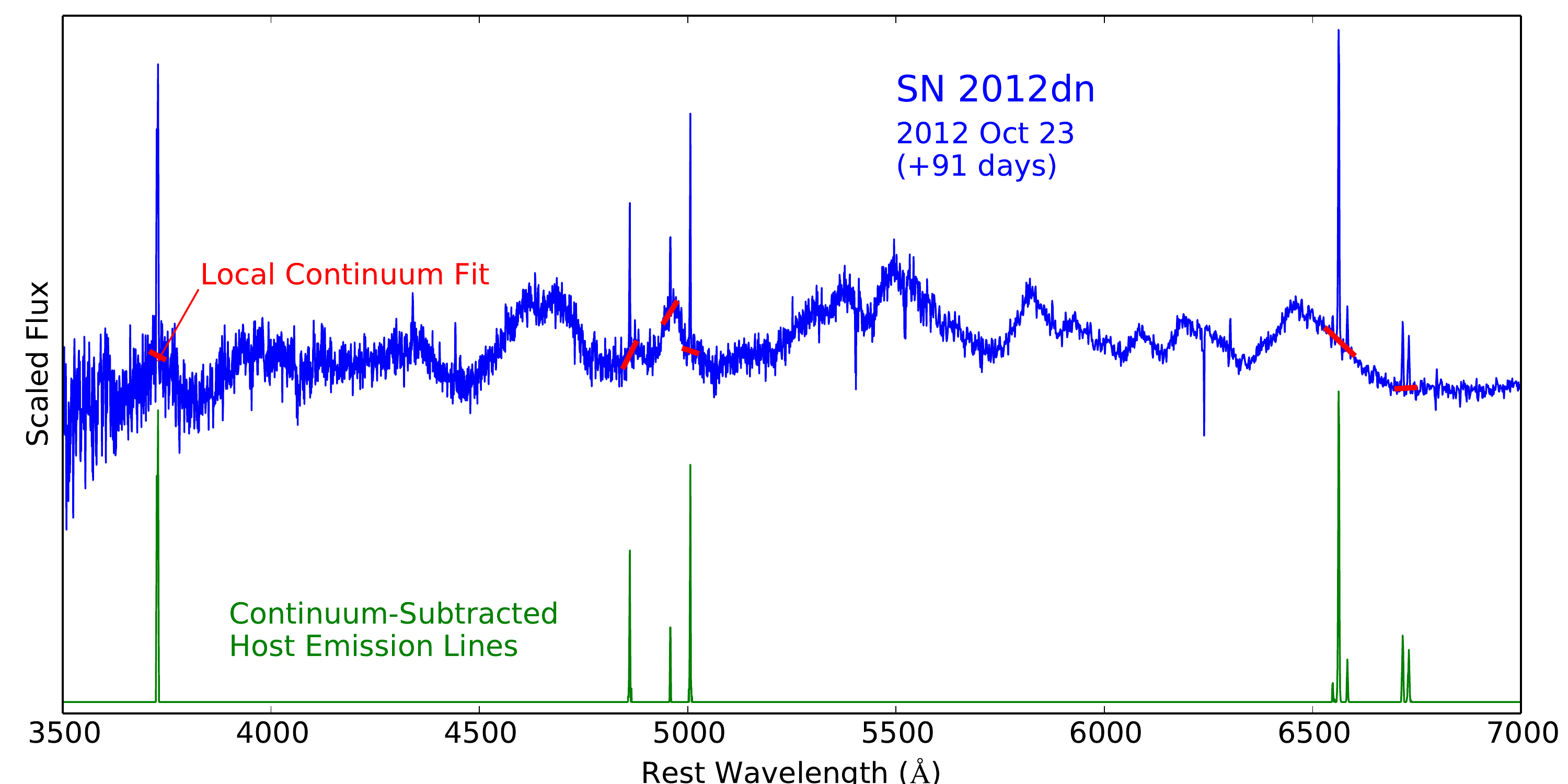}
\caption{Extraction of host galaxy emission line flux (green) from late SN~2012dn spectrum (blue) using simple linear continuum fits (red).}
\label{fig:sn2012dn_host_lines}
\end{center}
\end{figure}

Finally, the high resolution of \wifes\ reveals narrow emission lines from the host galaxy of SN~2012dn.  We isolated the narrow host galaxy lines using the longest exposure of SN~2012dn, the late-phase observation of 2012 Oct 23 (+91 days).  We fit a simple linear ``continuum'' near each emission line by fitting a line to the SN spectrum between 5 and 15 \AA\ away from the line centre on both the blue and red sides of the line.  We illustrate this technique and present the SN-subtracted emission features in Figure~\ref{fig:sn2012dn_host_lines}.

\begin{table}[h]
\caption{SN~2012dn Local Emission Line Fluxes}
\begin{tabular}{ l r r }
\hline
Line     &
$F(\lambda)/F(H\beta)$ & 
$F(\lambda)/F(H\beta)$ \\
& (raw)$^a$ & (de-reddened)$^b$ \\
\hline
$[$O\,\textsc{ii}$]$~$\lambda\lambda 3727,3730$ & $3.61 \pm 0.24$ & $4.33 \pm 0.29$ \\
H$\beta$                       & $1.00 \pm 0.10$ & $1.00 \pm 0.10$ \\
$[$O\,\textsc{iii}$]$~$\lambda 4959$            & $0.45 \pm 0.09$ & $0.44 \pm 0.09$ \\
$[$O\,\textsc{iii}$]$~$\lambda 5007$            & $1.64 \pm 0.10$ & $1.60 \pm 0.10$ \\
H$\alpha$                      & $3.40 \pm 0.07$ & $2.87 \pm 0.06$ \\
$[$N\,\textsc{ii}$]$~$\lambda 6548$             & $0.06 \pm 0.06$ & $0.05 \pm 0.05$ \\
$[$N\,\textsc{ii}$]$~$\lambda 6584$             & $0.37 \pm 0.06$ & $0.31 \pm 0.05$ \\
$[$S\,\textsc{ii}$]$~$\lambda 6717$             & $0.71 \pm 0.06$ & $0.59 \pm 0.05$ \\
$[$S\,\textsc{ii}$]$~$\lambda 6731$             & $0.58 \pm 0.06$ & $0.48 \pm 0.05$ \\
\hline
\end{tabular}
\\
$^a$ Observer frame fluxes, scaled to H$\beta$.\\
$^b$ Dereddened using Balmer decrement reddening of $E(B-V)=0.17$ so that $F(H\beta) = F(H\alpha)/2.87$ with a CCM reddening law, and scaled to the de-reddened flux of H$\beta$.
\label{tab:emlines}
\end{table}

In Table~\ref{tab:emlines} we present the measured emission fluxes and errors for the major galaxy emission lines, all of which have been scaled by the observed flux in the H$\beta$ line.  By comparing the observed ratio of the H$\alpha$ and H$\beta$ lines to its expected value of 2.87 \citep[the Balmer decrement][]{agn2}, we can determine the amount of reddening in the H\,\textsc{ii} regions giving rise to the host emission lines.  We use the \citet{ccm} to find a host reddening of $E(B-V)=0.17 \pm 0.10$, a value remarkably similar to the SN reddening of $E(B-V)=0.18$ (Milky Way plus host) determined by \citet{chakradhari14}.  We correct the host emission line fluxes for the value $E(B-V)=0.17$ and report the corrected values (which we also re-scale to the dereddened H$\beta$ flux) in Table~\ref{tab:emlines}.

With the de-reddened host galaxy emission line fluxes, we calculate a gas-phase metallicity at the site of SN~2012dn.  The N2 method of \citet[][hereafter PP04]{pp04} yields $12 + \log(O/H) = 8.29 \pm 0.04$, while the O3N2 method of PP04 yields $12 + \log(O/H) = 8.35 \pm 0.03$.  The former value is remarkably close to the estimated site metallicity for SN~2006gz of $12 + \log(O/H) = 8.26$ calculated by \citet{khan11} using the same metallicity method.  If we convert the O3N2 value to the \citet{trem04} scale using the formulae of \cite{ke08} as was done in \citet{childress11}, we measure $12 + \log(O/H)_{T04} = 8.51 \pm 0.04$.

Comparing the above values to the solar oxygen abundance of $12 + \log(O/H)_\odot = 8.69$ \citep{asplund09} we find the site metallicity for SN~2012dn is in the range 40-65\% solar, depending on the chosen metallicity calibration.  This sub-solar metallicity value is consistent with the previously reported trend for super-Chandra \sneia\ to prefer low metallicity environments \citep{taubenberger11, childress11, khan11}.

\section{Conclusions}
\label{sec:conclusions}
This work marks the primary data release for the ANU WiFeS SuperNovA Program (AWSNAP), comprising \nspec\ distinct spectra of \nobj\ unique supernovae.  These data were collected using the Wide Field Spectrograph (WiFeS) on the ANU 2.3m telescope during \nnights\ nights of observing over a three-year period from mid-2012 to mid-2015.

The AWSNAP spectroscopy sample is comparable in size to other SN spectra data releases, and its composition of SN types is roughly in line with expectations for a magnitude-limited SN search.  The phase coverage of the AWSNAP \snia\ sample is comparable to other published \snia\ spectroscopy datasets (for \sneia\ with multiple epochs of observation), with the inclusion of more \sneia\ with a single observation (i.e., classification spectra only).

We presented some analyses of the AWSNAP \snia\ sample, including some results uniquely enabled by the fine wavelength resolution available with \wifes.  We measured broad absorption features in \snia\ spectra at maximum light, including the ratio of silicon absorption features $R_{Si}$ and the strength of high velocity features $R_{HVF}$.  Additionally, we measured the strength and velocity of narrow sodium absorption features, including some cases where the integral-field nature of the instrument allowed us to measure the local systemic velocity within the SN host galaxy.  Some expected feature trends, such as a correlation between $R_{HVF}$ and $R_{Si}$, were recovered in our data set.  The nature of sodium absorption in our sample was limited by small number statistics.  Finally, we presented our observations of the candidate super-Chandrasekhar \snia\ SN~2012dn, and used narrow host galaxy emission features to show the SN site exhibits sub-solar metallicity.

The \wifes\ instrument presents several unique advantages for the study of transients, particularly owing to its comparatively narrow velocity resolution ($\sigma_v \sim 45$~\kms).  It has previously been employed in the study of SNe with strong narrow emission features such as SN~2009ip \citep{fraser13, fraser15}, SN~2012ca \citep{inserra14, inserra16}, and SN~2013fc \citep{kangas16}.  Here we also demonstrated that the fine velocity resolution allows for the measurement of narrow {\em absorption} features, particularly sodium absorption in \sneia.  The higher resolution of \wifes\ also frequently revealed narrow host galaxy emission features at the site of the SN, which can at times be used to determine a SN site metallicity (as we showed for SN~2012dn).  Finally, the integral field nature of \wifes\ allowed us to measure the local host galaxy rotational velocity at the site of several SNe, even when there was no emission directly at the SN location.  Thus \wifes\ is an instrument well-suited for not only standard SN spectroscopic observations, but also a unique suite of capabilities not commonly found in transient followup instruments.

\section*{Acknowledgements}
We are very grateful for the excellent technical support staff for the ANU 2.3m telescope and the \wifes\ instrument: Peter Verwayen, Ian Adams, Peter Small, Ian Price, Peter Young, and Jon Nielsen. We thank the ANU telescope time allocation committee who continue to support the observations presented herein.  We also thank Julie Banfield, Michael Ireland, Stefan Keller, Lisa Kewley, Jeremy Mould, Chris Owen, Aaron Rizzuto, Dary Ruiz, and Tian-Tian Yuan for additional observations.

This research was conducted by the Australian Research Council Centre of Excellence for All-sky Astrophysics (CAASTRO), through project number CE110001020.  IRS was supported by Australian Research Council Laureate Grant FL0992131.
This research has made use of the NASA/IPAC Extragalactic Database (NED) which is operated by the Jet Propulsion Laboratory, California Institute of Technology, under contract with the National Aeronautics and Space Administration.
This research has made use of NASA's Astrophysics Data System (ADS).

\bibliographystyle{apj}
\bibliography{awsnap_DR1_paper.bib}

\section*{Appendix A1: Target and Observation Details}
\label{app:details}
In this Appendix we present the observational metadata for the AWSNAP data release, as well as sodium fit results and host (or local) redshifts used in the sodium fitting.

{\em AWSNAP targets:}
Table~\ref{tab:awsnap_targets} presents the discovery and classification details for the full list of \nobj\ objects featured in the AWSNAP DR1 data release.  We list the discovery location and date, as well as discovery reference and group that discovered the transient.  Note the discovery references are typically either Astronomer's Telegrams (ATel's, denoted as, e.g., ``A1234'') or telegrams from the Central Bureaur for Electronic Telegrams (CBET's, denoted as, e.g., ``C1234''), though occasionally a separate discovery notice was not issued -- in these cases typically the webpage of the discovering group was cited in the classification reference.  Classification references are given with similar notation, listed along with with classification date, quoted supernova type and phase from the classification, as well as the redshift listed in the classification.

\begin{landscape}
\begin{deluxetable}{lllllllllllc}
\tablewidth{0pt}
\tablecaption{All AWSNAP Targets
\label{tab:awsnap_targets}}
\tabletypesize{\small}
\tablehead{
\colhead{SN} & 
\multicolumn{5}{c}{Discovery} &
\multicolumn{6}{c}{Classification}\\
\colhead{} & 
\colhead{RA} & 
\colhead{DEC} & 
\colhead{Ref.$^a$} & 
\colhead{Date} & 
\colhead{Group$^b$} & 
\colhead{Ref.} & 
\colhead{Date} & 
\colhead{Group$^c$} & 
\colhead{SN Type} & 
\colhead{Redshift} & 
\colhead{Phase$^d$}
}
\startdata
SN~2009ip             &  22:23:08.1  &  -28:56:35  &  A4334  &  20120724  &  CRTS        &  A4338   &  20120826  &  Foley            &  SNIIn      &  0.005944  &  (unk)   \\
SN~2012ca             &  18:41:06.5  &  -41:47:38  &  C3101  &  20120426  &  Parker      &  A4076   &  20120429  &  PESSTO           &  SNIIn      &  0.019000  &  (unk)   \\
SN~2012dj             &  23:14:47.2  &  -43:36:22  &  C3167  &  20120702  &  Parker      &  C3167   &  20120706  &  Parrent          &  SNIb/c     &  0.005324  &  +0      \\
SN~2012dn             &  20:23:36.3  &  -28:16:43  &  C3174  &  20120708  &  Parker      &  A4253   &  20120712  &  SNF              &  SNIa       &  0.010187  &  -10     \\
SN~2012dt             &  00:56:38.1  &  -09:53:59  &  C3188  &  20120718  &  Parker      &  A4269   &  20120721  &  PTF              &  SNII       &  0.018900  &  +9      \\
SN~2012dy             &  21:18:49.5  &  -57:38:42  &  C3197  &  20120804  &  Bock        &  C3197   &  20120805  &  Milisavljevic    &  SNII       &  0.010300  &  (young) \\
SN~2012ec             &  02:45:59.9  &  -07:34:15  &  C3201  &  20120811  &  Monard      &  A4306   &  20120812  &  AWSNAP           &  SNII       &  0.004693  &  (young) \\
SN~2012eq             &  01:00:14.5  &  -30:48:26  &  C3223  &  20120827  &  CRTS        &  A4362   &  20120907  &  PESSTO           &  SNIa       &  0.032230  &  +45     \\
SN~2012eu             &  03:13:04.2  &  -08:23:24  &  C3231  &  20120827  &  CRTS        &  C3231   &  20120904  &  Yang             &  SNIa       &  0.030000  &  +30     \\
SN~2012fr             &  03:33:37.1  &  -36:07:28  &  A4523  &  20121027  &  TAROT       &  A4525   &  20121028  &  AWSNAP           &  SNIa       &  0.005400  &  -11     \\
\enddata
\tablecomments{
Abbreviated version -- full version of this table included in arXiv package. \\
$^a$ Notation: A1234 $\equiv$ ATel \# 1234; C1234 $\equiv$ CBET \# 1234; CTOCP $\equiv$ CBET Transient Object Confirmation Page; objects newly classified here denoted as (NEW); all others refer to transient announcement webpages for stated discovery group(s).\\
$^b$ Discovery group acronyms (where appropriate -- see text for details) or lead author of the discovery alert.\\
$^c$ Classifcation group acronyms (where appropriate -- see text for details) or lead author of the classification alert.\\
$^d$ Phase with respect to maximum light (where appropriate) for the best {\em spectroscopic match}, as reported by the classifying group.  SNe IIn do not typically have a clearly identifiable spectroscopic phase so are listed as ``(unk)'' (i.e., unknown).  Very young SNe II (with very blue featureless continua) similarly have ambiguous phases so are listed as ``(young)''.
}
\end{deluxetable}
\end{landscape}

In Table~\ref{tab:awsnap_targets} the discovery and classification groups are listed either as acronyms of the group name, or listed as the first author in the discovery/classification notice.  The acronyms used in the table for discovery groups are:
\begin{itemize}
  \item ASAS-SN: the All-Sky Automated Survey for Supernovae \citep{asassn}
  \item CHASE: the CHilean Automatic Supernova sEarch \citep{chase}
  \item CRTS: the Catalina Real-Time Transient Survey \citep{crts}
  \item Gaia: transient alerts from Gaia \citep[e.g.,][]{wyrz12, blago14, blago16}
  \item ISSP: the Italian Supernova Search Program
  \item LOSS: the Lick Observatory Supernova Search \citep{ganesh10}
  \item LSQ: the La Silla-QUEST low redshift supernova survey \citep{lsq}
  \item MASTER: the Mobile Astronomical System of Telescope-Robots \citep{master}
  \item OGLE: the OGLE-IV real-time transient search \citep{ogle}
  \item PanSTARRS: the Panoramic Survey Telescope and Rapid Response System {panstarrs}
  \item PTF/iPTF: the Palomar Transient Factory \citep{rau09, law09} and its successor
  \item SkyMapper: the SkyMapper \citep{keller07} transients search
  \item TAROT: the T\'elescope \'a Action Rapide pour les Objets Transitoires (Rapid Action Telescope for Transient Objects) at La Silla \citep[e.g.,][]{klotz13}
  \item TNTS: the Tsinghua-NAOC Transient Survey
\end{itemize}
The acronyms used in the table for classification groups are:
\begin{itemize}
  \item ASP: the Asiago Supernova Program \citep{tomasella14}
  \item AWSNAP: the ANU+WiFeS SuperNovA Program (this work)
  \item CSP: the Carnegie Supernova Project \citep{folatelli13}
  \item LCOGT: the Las Cumbres Observatory Global Telescope \citep{lcogt}
  \item PESSTO: the Public ESO Spectroscopic Survey for Transient Objects \citep{pessto}
  \item SNF: SuperNova Factory \citep{aldering02}
\end{itemize}

Some transients in Table~\ref{tab:awsnap_targets} are known by other aliases or are listed in shorthand form (due to a lengthy full name).  These transients and their aliases (or full names) are noted in Table~\ref{tab:awsnap_aliases}.

\begin{table}
\begin{center}
\caption{Alternate and/or full designations for SNe in the AWSNAP sample}
\label{tab:awsnap_aliases}
\scriptsize
\begin{tabular}{ll}
\hline
SN      & Alias \\
\hline
SN~2013ef             & ASASSN-13bb   \\
SN~2014E              & PTF14w        \\
SN~2014cx             & ASASSN-14gm   \\
SN~2014dq             & ASASSN-14jb   \\
SN~2015L              & ASASSN-15lh   \\
iPTF13dge             & PS1-13dvn     \\
LSQ15adm              & Gaia15aep     \\
LSQ15ey               & PS15ii        \\
PS15ae                & CSS141223:113342+004332 \\
PSNJ09023787+2556042  & Gaia15aet     \\
PSNJ10433393-3048206  & PS15ip        \\
PSNJ20250386-2449133  & PS15bjv       \\
SNhunt222             & LSQ14rl       \\
CSS1222               & PS1-14ya      \\
CSS1222               & CSS140326:122257+282955 \\
MASTERJ1408           &  MASTER OT J140804.26-115949.7 \\
MASTERJ1353           &  MASTER OT J135329.90-421622.5 \\
MASTERJ0746           &  MASTER OT J074610.09-712224.9 \\
\hline
\end{tabular}
\normalsize
\end{center}
\end{table}

{\em AWSNAP spectra:}
The full list of AWSNAP spectra released here is given in Table~\ref{tab:all_awsnap_spectra}, along with pertinent observation details such as grating and exposure time.

\begin{deluxetable}{lccr}
\tablewidth{0pt}
\tablecaption{All AWSNAP Spectra
\label{tab:all_awsnap_spectra}}
\tabletypesize{\small}
\tablehead{
\colhead{SN} & 
\colhead{Observation} & 
\colhead{Grating} & 
\colhead{Exposure} \\
\colhead{} & 
\colhead{Date} & 
\colhead{} & 
\colhead{Time (s)}
}
\startdata
\scriptsize
SN~2009ip             &  2012-Oct-22  &  B3000  &   1200  \\
SN~2009ip             &  2012-Oct-22  &  R3000  &   1200  \\
SN~2009ip             &  2012-Oct-23  &  B7000  &   1200  \\
SN~2009ip             &  2012-Oct-23  &  R7000  &   1200  \\
SN~2009ip             &  2012-Sep-22  &  B3000  &   1200  \\
SN~2009ip             &  2012-Sep-22  &  R3000  &   1200  \\
SN~2009ip             &  2012-Sep-23  &  B3000  &   1200  \\
SN~2009ip             &  2012-Sep-23  &  R3000  &   1200  \\
SN~2009ip             &  2013-Apr-12  &  B7000  &   1200  \\
SN~2009ip             &  2013-Apr-12  &  R7000  &   1200  \\
\normalsize
\enddata
\tablecomments{
Abbreviated version -- full version of this table included in arXiv package.
}
\end{deluxetable}

{\em SN~Ia sodium fits:}
In Table~\ref{tab:nad_fit_results} we present the best fit parameters (and uncertainties) for our fits to narrow sodium absorption in our sample of \snia\ maximum light spectra.  Table~\ref{tab:nad_host_redshifts} lists the redshifts used to establish the local rest velocity of the SN -- typically this was the redshift of the host galaxy but occasionally is obtained from galaxy emission features at the location of the SN (denoted as ``Local'').

\begin{center}
\begin{table*}
\caption{Fits of Na line in SNe~Ia}
\label{tab:nad_fit_results}
\begin{tabular}{lccrrr}
\hline
SN  &  Observation  &  Grating  &  $v_{Na}$  &  $\sigma_{Na}$  &  $pEW_{Na}$  \\
    &  Date         &           &  (\kms)   &   (\kms)    &  (\AA)  \\
\hline
ASASSN-14hr           &  2014-Oct-03 &  R3000  &  $-41.3 \pm 16.5$  &  $ 76.3 \pm 13.6$  &  $0.74 \pm 0.17$  \\
ASASSN-14jc           &  2014-Oct-27 &  R3000  &  $ -2.8 \pm 11.5$  &  $ 65.3 \pm 10.6$  &  $0.98 \pm 0.19$  \\
ASASSN-14jg           &  2014-Nov-06 &  R3000  &  $  8.9 \pm 15.3$  &  $ 58.1 \pm 15.2$  &  $0.23 \pm 0.07$  \\
ASASSN-15go           &  2015-Apr-11 &  R3000  &  $ 82.9 \pm 12.1$  &  $ 70.0 \pm 10.6$  &  $1.75 \pm 0.32$  \\
iPTF13dge             &  2013-Sep-19 &  R7000  &  $-35.1 \pm  3.3$  &  $ 22.3 \pm  3.4$  &  $0.34 \pm 0.05$  \\
LSQ14bcj              &  2014-Apr-23 &  R7000  &  $ 20.9 \pm  4.7$  &  $ 28.0 \pm  4.9$  &  $0.99 \pm 0.19$  \\
PS1-14wl              &  2014-Apr-02 &  R3000  &  $ 35.9 \pm 12.9$  &  $ 70.0 \pm  0.0$  &  $1.22 \pm 0.22$  \\
PSNJ13471211-2422171  &  2015-Feb-15 &  R3000  &  $ 14.5 \pm 49.6$  &  $119.4 \pm 35.9$  &  $0.71 \pm 0.26$  \\
PSNJ17194328-7721305  &  2013-Aug-30 &  R7000  &  $-70.7 \pm  3.2$  &  $ 24.3 \pm  3.4$  &  $0.47 \pm 0.07$  \\
SN~2012dn             &  2012-Jul-21 &  R3000  &  $ -9.4 \pm 19.8$  &  $ 79.0 \pm 17.2$  &  $0.25 \pm 0.07$  \\
SN~2013aj             &  2013-Mar-08 &  R3000  &  $  5.0 \pm 24.5$  &  $ 95.2 \pm 18.8$  &  $0.43 \pm 0.11$  \\
SN~2013cg             &  2013-May-26 &  R7000  &  $ 42.0 \pm 23.2$  &  $118.9 \pm 16.6$  &  $1.39 \pm 0.25$  \\
SN~2013cy             &  2013-Jun-03 &  R3000  &  $ 45.2 \pm 19.0$  &  $ 42.5 \pm 20.8$  &  $0.19 \pm 0.11$  \\
SN~2013er             &  2013-Jul-31 &  R3000  &  $ 11.4 \pm 16.3$  &  $ 78.4 \pm 13.3$  &  $2.05 \pm 0.44$  \\
SN~2013gh             &  2013-Aug-30 &  R7000  &  $-18.9 \pm  3.0$  &  $ 23.3 \pm  3.1$  &  $1.50 \pm 0.22$  \\
SN~2014aa             &  2014-Mar-14 &  R3000  &  $ -4.7 \pm 43.5$  &  $130.0 \pm  0.0$  &  $1.42 \pm 0.34$  \\
SN~2014ao             &  2014-Apr-23 &  R7000  &  $ -7.2 \pm  3.1$  &  $ 23.7 \pm  3.2$  &  $1.33 \pm 0.19$  \\
SN~2014at             &  2014-Apr-23 &  R3000  &  $  0.3 \pm 41.8$  &  $130.0 \pm  0.0$  &  $0.40 \pm 0.14$  \\
SN~2014bx             &  2014-Jul-22 &  R3000  &  $ 55.4 \pm 12.9$  &  $ 73.7 \pm 10.9$  &  $4.34 \pm 0.79$  \\
\hline
\end{tabular}
\end{table*}
\end{center}

\begin{center}
\begin{table*}
\caption{Host Galaxy Redshifts for SN~Ia Na sample}
\label{tab:nad_host_redshifts}
\begin{tabular}{llll}
\hline
SN  &  Host  &  Redshift  & Ref.$^a$  \\
\hline
ASASSN-14hr           &  2MASX~J01504127-1431032  &  0.033620  &  6dF (Jones)  \\
ASASSN-14jc           &  2MASX J07353554-6246099  &  0.011320  &  6dF (Jones)  \\
ASASSN-14jg           &  PGC~128348               &  0.015315  &  Local  \\
ASASSN-15go           &  (unknown)                &  0.019030  &  Local  \\
iPTF13dge             &  NGC~1762                 &  0.015854  &  \citep{theureau98}  \\
LSQ14bcj              &  MCG~-02-36-015           &  0.036056  &  Local  \\
PS1-14wl              &  2MASX J13324055-2036204  &  0.032346  &  Local  \\
PSNJ13471211-2422171  &  ESO~509-G108             &  0.019910  &  \citep{ogando08}  \\
PSNJ17194328-7721305  &  ESO~044-G010             &  0.009587  &  Local  \\
SN~2012dn             &  ESO~462-G016             &  0.010187  &  \citep{theureau98}  \\
SN~2013aj             &  NGC~5339                 &  0.009126  &  \citep{theureau98}  \\
SN~2013cg             &  NGC~2891                 &  0.007952  &  \citep{ogando08}  \\
SN~2013cy             &  ESO~532-G025             &  0.031465  &  \citep{dacosta91}  \\
SN~2013er             &  IC~850                   &  0.018052  &  \citep{grogin98}  \\
SN~2013gh             &  NGC~7183                 &  0.009075  &  Local  \\
SN~2014aa             &  NGC~3861                 &  0.016982  &  \citep{cortese08}  \\
SN~2014ao             &  NGC~2615                 &  0.014663  &  Local  \\
SN~2014at             &  NGC~7119                 &  0.032242  &  \citep{donzelli00}  \\
SN~2014bx             &  NGC~6808                 &  0.011570  &  \citep{rsa81}  \\
\hline
\end{tabular}
\\
$^a$ Via NED.  ``Local'' denotes local velocity measured with WiFeS, either from narrow lines in the SN spectrum or from galaxy rotation characteristics (see text for details).
\end{table*}
\end{center}

\end{document}